%% file: CPFA.tex
\documentclass[11pt]{article}
\usepackage{amsmath, amssymb, amsthm,authblk}
\usepackage{natbib}
\usepackage{epsfig,euscript}
 \usepackage{mathabx,multirow}
\usepackage{latexsym, lscape}
 \usepackage{subfigure}
 \usepackage{mathtools}
 \usepackage{enumitem}
 \usepackage{tikz,tabularx}
\renewcommand{\baselinestretch}{1.5}
\makeatletter
\def\endp{{\hfill \vrule width 5pt height 5pt\par}}

\newcommand{\cM}{{\cal M}}
\newcommand{\rE}{{\rm E}}

\newcommand{\cD}{\mbox{$\mathcal{D}$}}

\input{alias}

\bibpunct{(}{)}{;}{a}{,}{,}

\title{\bf Factor Analysis for High-Dimensional Time Series with Change Point\footnote{Xialu Liu is Assistant Professor, Management Information Systems Department, San Diego State University, San Diego, CA 92182. Email: xialu.liu@sdsu.edu.
Ting Zhang is Assistant Professor, Department of Mathematics and Statistics, Boston University, Boston, MA 02215. E-mail: tingz@bu.edu. Xialu Liu is the corresponding
author.}}
\author{Xialu Liu}
\affil{San Diego State University}
\author{Ting Zhang}
\affil{Boston University}
\begin{document}
\maketitle
\date{}

\begin{abstract}
We consider change-point latent factor models for high-dimensional time series, where a structural break may exist in the underlying factor structure. In particular, we propose consistent estimators for factor loading spaces before and after the change point, and the problem of estimating the change-point location is also considered. Compared with existing results on change-point factor analysis of high-dimensional time series, a distinguished feature of the current paper is that our results allow strong cross-sectional dependence in the noise process. {To accommodate the unknown degree of cross-sectional dependence strength}, we propose to use self-normalization to pivotalize the change-point test statistic. Numerical experiments including a Monte Carlo simulation study and a real data application are presented to illustrate the proposed methods.

\end{abstract}

\section{Introduction}\label{sec:introduction}
High-dimensional time series has been emerging as a common and important data type in applications from a number of disciplines, including climate science, economics, finance, medical science, and telecommunication engineering among others. Although numerous statistical methods and their associated theory have been developed for the modeling and inference of time series data, existing results mostly focused on the univariate or finite-dimensional multivariate case. The problem of extending existing results developed under low-dimensional settings to handle high-dimensional time series, however, is typically nontrivial and requires significant innovations. For example, when the dimension is larger than the length of the observed time series, the commonly used autoregressive moving-average (ARMA) model in its conventional form may face a serious identification problem as commented by \citet{lam2011}. To handle the phenomenon of high dimensionality, one typically resorts to certain sparsity-type conditions for the purpose of dimension reduction. For example, when considering vector autoregressive (VAR) models in the high-dimensional setting, one typically need to assume that the coefficient matrices are sparse in a suitable sense in order to obtain their meaningful estimators; see for example \citet{Basu:Michailidis:2015}, \citet{Davis:Zang:Zhen:2016} and references therein for research in this direction.

Unlike the aforementioned sparse VAR approach that aims at extending existing parametric time series models to their sparse high-dimensional counterparts, a more commonly used approach in the literature for modeling high-dimensional time series is through the use of factor models; see for example \citet{chamberlain1983} and \citet{stock1998}, \citet{bai2002}, and \citet{forni2004} among others. {In the aforementioned works, it was assumed that most of the variations in the observed high-dimensional time series can be explained by a set of common factors, and as a result such factor models cannot be used to capture strong cross-sectional dependence. In addition, the common component  suffers from the identifiability issue when the dimension is finite; see for example \citet{bai2002}. To alleviate these problems, \citet{lam2011} proposed an alternative type of factor models which has become more and more popular in the last decade. In the model of \citet{lam2011}, the common factors are viewed as the force that drives all the dynamics and are used to explain the serial dependence in the data. Under this setting, the noise process can accommodate the strong cross-sectional dependence and is white. \citet{lam2011} proposed an approach based on autocovariance matrices of the observed process at nonzero lags for loading space estimation. This method is also applicable to non-stationary processes, processes with uncorrelated or endogenous regressors, and matrix-valued process; see for example \citet{liu2016}, \citet{chang2017},  \citet{wang2019}, and \citet{liu}. Compared with the approach of \citet{bai2002} and \citet{forni2004}, the factor model by \citet{lam2011} not only ensures that the common component is identifiable but also captures all serial dependence of data which enables us to build forecasting models after dimension reduction. However, existing results along the line of \citet{lam2011} were generally developed under the assumption that the underlying factor structure remains the same over the whole time period, while recent empirical applications reveal that the factor structure tends to exhibit structural breaks at a certain point, either due to a sudden market change or in response to some unpredictable reasons. This motivates us to consider factor models with possible change points.}

The problem of change-point analysis for factor models has been an active area of research. For this, \citet{breitung2011} considered testing the existence of change in loadings when the noise process is either independent or autoregressive with a finite order. \citet{Chen:Dolado:Gonzalo:2014} came up with a Lagrange multiplier test and a Wald test for detecting the break by regressing one of the factors estimated by PCA on the remaining estimated factors. \citet{han2015} testes the break point through the second moments of the estimated factors. \citet{barigozzi2018} constructed a cumulative sum (CUSUM) test using wavelet coefficients for detecting multiple structural breaks when the noise sequence is Gaussian. Besides testing-based methods, \citet{chen2015} and \citet{baltagi2017} considered using methods based on least squares to estimate the change-point location. \citet{Ma:Su:2018} proposed an adaptive fused group lasso approach to estimate the multiple change-point location. However, existing results in this direction were typically developed for models with the so-called idiosyncratic noise, where no strong cross-sectional dependence is allowed in the noise. In the current paper, we follow settings by \citet{lam2011}  where the common factor drives all the dynamics and explain serial dependence of the observed process, and consider the situation when the strength of cross-sectional dependence of the noise sequence is unknown and potentially strong. {As discussed above, one advantage of this setting is that we can fully extract the dependence of data through common factors for future forecasting model building if needed.} The aforementioned paper considered the problem of estimating the factor loading by assuming that there is no change point, and we shall here focus on the change-point case.

The remaining of the paper is organized as follows. Section \ref{sec:estimation} introduces the change-point factor model and considers its associated estimation problem, including estimating factor loading spaces before and after the change point and the location of the change point. The asymptotic properties of the proposed estimators are also investigated. Section \ref{subsec:cptest} proposes a self-normalized approach to testing the existence of change point, which has a pivotalized asymptotic distribution regardless of whether the noise process has weak or strong cross-sectional dependence. {Details on its practical implementation are discussed in Section \ref{subsec:teststat}. Numerical experiments, including a Monte Carlo simulation study and a real data application, are presented in Sections \ref{sec:numericalexperiments} and \ref{sec:realdata} respectively to illustrate the proposed methods. Section \ref{sec:conclusion} concludes the paper. Technical proofs are deferred to the Appendix.}

\section{Change-Point Factor Model and Its Estimation}\label{sec:estimation}
{In this section, we will introduce the our change-point factor model and study its associated estimation problem, including the loading space before and after the change point and the change point location.}
\subsection{Change-Point Factor Model}\label{subsec:cpmodel}
Suppose we observe a $p$-dimensional time series $\by_t$, $t = 1,\ldots,n$, according to a factor model, then one can write
\begin{equation}\label{lam}
\by_t=\bA \bx_t +\bve_{t},
\end{equation}
where $(\bx_t)$ is a latent factor process whose dimension $k_0$ is typically much smaller than $p$, $\bA \in \mathbb R^{p \times k_0}$ is the associated loading matrix, and $(\bve_t)$ denotes the white noise process. The latent factor model (\ref{lam}) has been widely used in the literature for dimension reduction of high-dimensional time series; see for example \citet{bai2002}, \citet{lam2011}, \citet{lam2012}, \citet{chang2015} and references therein. It also relates to the generalized dynamic factor model of \citet{forni2005} in which the latent factor process $(\bx_t)$ is assumed to follow a low-dimensional autoregressive model. We shall here consider the general setting where $(\bx_t)$ is not necessarily an autoregressive process. In model (\ref{lam}), the factor loading structure remains the same over the whole sampling period, while in many applications a structural break may occur due to various reasons. For this, we consider the change-point factor model
\begin{equation}\label{model}
\by_t = \left\{\begin{array}{ll}
\bA_1\bx_{t,1} + \bve_t, & \mathrm{if}\ t \leq r_0; \\
\bA_2\bx_{t,2} + \bve_t, & \mathrm{if}\ t > r_0,
\end{array}\right.
\end{equation}
{where $\bx_{t,i}\in \mathbb R^{k_i}$, $i = 1,2$, represents the underlying latent factor before and after the change point whose location is denoted by $r_0$, and $\bA_1$ and $\bA_2$ are the associated loading matrices.} Recently, there have been efforts in studying the change-point factor model by incorporating certain beliefs on the change-point mechanism into the analysis. For example, \citet{liu2016} modeled the change-point mechanism by a finite-state hidden Markov chain, and thus structural breaks occur when there is a regime switching in the hidden state variable. On the other hand, \citet{liu} considered using a threshold variable to model the change-point mechanism, where the threshold variable is assumed to be $\alpha$-mixing and observable up to a small number of unknown parameters. Instead of introducing a Markov chain process or an additional threshold variable, we in the current paper focus on the change-point factor model (\ref{model}) which uses time to naturally divide the observed process into {homogenous} pieces before and after the change point.

In practice, one does not observe $(\bx_{t,i})$ nor $(\bve_t)$ but only $(\by_t)$, and thus the loading matrices in (\ref{model}) are not fully identifiable. To be more specific, for $i = 1,2$, one can replace $(\bA_i, \bx_{t,i})$ in (\ref{model}) by $(\bA_i \bU_i,\bU^{-1}_i \bx_t)$ for any $k_i \times k_i$ non-singular matrix $\bU_i$. However, the space spanned by the columns of the loading matrix $\bA_i$, denoted by ${\cal M}(\bA_i)$, is always uniquely defined; see for example the discussions in \cite{lam2011}, \citet{chang2015}, \citet{liu2016} and \citet{liu}. As a result, when estimating the factor model (\ref{model}), we shall focus on the loading space ${\cal M}(\bA_i)$ instead of the loading matrix $\bA_i$ itself. 

We shall here introduce some notations. For a matrix $\bH$, we use $\|\bH\|_F$ and $\|\bH\|_2$ to denote its Frobenius and L-2 norms respectively. In addition, we use ${\rm tr}(\bH)$ for the trace, $\sigma_i(\bH)$ for the $i$-th largest singular value, and $\|\bH\|_{\min}$ for the square root of minimum nonzero eigenvalue of $\bH'\bH$. Also, we write $a \asymp b$ if $a=O(b)$ and $b=O(a)$, and we use $\lfloor{x}\rfloor$ and $\lceil{x}\rceil$ to denote the largest previous and smallest following integers of $x$. 

\subsection{Estimating the Loading Spaces}\label{subsec:estimation}
{Before introducing our estimators for loading spaces in the change-point setting, we first need a measure to quantify the distance between two linear spaces which can then be used to assess the statistical performance our estimators. For this, let $\bS_1$ and $\bS_2$ be full rank matrices in $\mathbb R^{p\times q_1}$ and $\mathbb R^{p\times q_2}$ respectively with $\max(q_1,q_2) \leq p$. Denote $\bO_i$ the matrix whose columns form an orthonormal basis of $\cM(\bS_i)$ for $i = 1,2$, then the distance between column spaces of $\bS_1$ and $\bS_2$ can be measured by
\begin{equation}\label{distance_diff}
{\cal D}\{\cM(\bS_1),\,\cM(\bS_2)\} = \left\{1 - \frac{{\rm tr}(\bO_1 \bO_1' \bO_2 \bO_2')}{\min(q_1,q_2)}  \right\}^{1/2}.
\end{equation}
The distance measure (\ref{distance_diff}) was first introduced in \cite{liu}, and is a quantity between 0 and 1. In particular, it equals to 0 if ${\cal M}(\bS_1) \in {\cal M}(\bS_2)$ or
${\cal M}(\bS_2) \in {\cal M}(\bS_1)$, and equals to 1 if $\cM(\bS_1)$ and $\cM(\bS_2)$ are orthogonal. For the special case when $q_1 = q_2 = q$, the two spaces $\bS_1$ and $\bS_2$ have the same dimension, and the distance measure (\ref{distance_diff}) reduces to
\begin{equation}\label{distance}
{\cal D}\{\cM(\bS_1),\,\cM(\bS_2)\}
= \left\{1 - \frac{{\rm tr}(\bO_1 \bO_1'
\bO_2 \bO_2')}{q}  \right\}^{1/2},
\end{equation}
which was used in \citet{chang2015} and \citet{liu2016}. Since the number of factors is usually unknown in practice and may be estimated in a nonperfect way, we shall in the current paper use the generalized version in (\ref{distance_diff}) to measure the distance between two linear spaces.}

For factor models in high-dimensional cases, it is common to assume that the number of factors is fixed but the squared L-2 norm of the $p \times k_i$ loading matrix $\bA_i$ grows with the dimension $p$ \citep{bai2002,doz2011}. The growth rate is called the
strength of the factors in \citet{lam2011}, \citet{lam2012}, \citet{chang2015}, and \citet{liu2019}. {For $i = 1,2$, assume that}
\begin{eqnarray*}
\|\bA_i\|_2^2 \asymp \|\bA_i\|_{\min}^2 \asymp p^{1-\delta_i}
\end{eqnarray*}
{for some $0\leq \delta_i \leq 1$, then factors in regime $i$ are said to be strong if $\delta_i = 0$ and weak if $\delta_i \in (0,1)$.} The strength of factors measures the relative growth rate of
the amount of information carried by
the observed process $\by_t$ about the
common factors $\bx_t$ as $p$ increases,
with respect to the growth rate of the amount of noise process in regime $i$. {It can be seen from our theoretical results below that the factor strength plays an important
role in the estimation efficiency.}

Let $\gamma_0=r_0/n$, {and we shall now} introduce the estimation procedure for loading spaces when a tentative break date $\gamma$ is given. {For this, we separate the data into two subsets before and after $\gamma$, namely $t \leq \lfloor \gamma n \rfloor$ and $t > \lfloor \gamma n \rfloor$, and propose a change-point generalization of the estimator of \citet{lam2011}. To be more specific, define the generalized second cross moment matrices of $\bx_{t,i}$ and $\by_t$ with lag $h$ in each regime $i$ as}
\begin{align*}
\bSigma_{x,i}(h,\gamma)=\frac{1}{n}\sum_{t=1}^{n}{\rm E} \left[\bx_{t,i} \bx_{t+h,i}' I_{t,i}(\gamma) I_{t+h,i}(\gamma) \right],\quad  \bSigma_{y,i}(h,\gamma)=\frac{1}{n}\sum_{t=1}^{n}{\rm E} \left[\by_t \by_{t+h}' I_{t,i}(\gamma) I_{t+h,i}(\gamma)\right],
\end{align*}
where $I_{t,i}(h, \gamma)$ is the indicator function for regime $i$ {satisfying} $I_{t,1}(\gamma)=1$ if $1 \leq t \leq \lfloor \gamma n \rfloor$, $I_{t,2}(h,\gamma)=1$ if $ \lfloor \gamma n \rfloor <t \leq n$, and zero otherwise. With the white noise assumption, $\bSigma_{x,i}(h,\gamma)$ and $\bSigma_{y,i}(h,\gamma)$ satisfy
\[
\bSigma_{y,1}(h,\gamma)= \bA_1 \bSigma_{x,1}(h,\gamma) \bA_1', \mbox{ if } \gamma \leq \gamma_0,
\]
and
\[
\bSigma_{y,2}(h,\gamma)= \bA_2 \bSigma_{x,2}(h,\gamma) \bA_2', \mbox{ if } {\gamma > \gamma_0}.
\]
For a pre-determined positive integer $h_0$, let
\begin{align}\label{M}
\bM_i(\gamma)=\sum_{h=1}^{h_0}\bSigma_{y,i}(h,\gamma) \bSigma_{y,i}(h,\gamma)',
\end{align}
{then we can see that}
\begin{equation}\label{Mi}
\bM_i(\gamma)=\sum_{h=1}^{h_0} \bA_i\left[ \bSigma_{x,i}(h,\gamma)\bA_i'\bA_i\bSigma_{x,i}(h,\gamma)'\right]\bA_i'
\end{equation}
{holds for $i=1$ if $\gamma \leq \gamma_0$ and for $i=2$ if $\gamma \geq \gamma_0$. Therefore, $\bM_i(\gamma)$ is a symmetric non-negative definite matrix sandwiched by $\bA_i$ and $\bA_i'$ when $\gamma$ is in regime $i$.} If there exists at least one nonzero $h$ such that $1\leq h \leq h_0$
and $\bSigma_{x,i}(h,\gamma)$ is full rank, then $\bM_i(\gamma)$ has rank $k_i$.
Its eigenspace corresponding to the nonzero eigenvalues is ${\cal M}(\bA_i)$.
Hence, ${\cal M} (\bA_i)$ can be estimated through an eigen-decomposition of the
sample version of $\bM_i$. {In particular}, let $\bq_{i,k}(\gamma)$  be the unit eigenvector of
$\bM_i(\gamma)$ corresponding to the $k$-th largest nonzero eigenvalue for $k=1,\ldots, k_i$.
Define
\begin{equation}
\bQ_i(\gamma)=(\bq_{i,1}(\gamma), \ldots, \bq_{i,k_i}(\gamma)),
\label{def:QB}
\end{equation}
where $k_i$ is the number of factors. For simplicity, in the rest of this paper we use $\bQ_i$ and $\bM_i$ to denote $\bQ_i(\gamma_0)$ and $\bM_i(\gamma_0)$ for $i=1,2$.

{To estimate the loading spaces in the change-point setting, we shall use the sample version of the above quantities and define
\[
\hat{\bSigma}_{y,i}(h,\gamma)= \frac{1}{n}\sum_{t=1}^{n} \left(\by_t \by_{t+h}' I_{t,i}(\gamma) I_{t+h,i}(\gamma) \right),\quad \hat{\bM}_i(\gamma)= \sum_{h=1}^{h_0}
\hat{\bSigma}_{y,i}(h,\gamma) \hat{ \bSigma}_{y,i}(h,\gamma)',
\]
for $i = 1,2$.} Let $\hat{\lambda}_{i,1}\geq \hat{\lambda}_{i,2}  \geq \ldots \geq  \hat{\lambda}_{i,p}$ be the $p$ eigenvalues of $\hat{\bM}_i(\gamma)$ and $\hat{\bq}_{i,1}(\gamma), \hat{\bq}_{i,2}(\gamma), \ldots, \hat{\bq}_{i,p}(\gamma)$ be the set of corresponding orthonormal
eigenvectors with $\mathbf{1}'\hat{\bq}_{i,j}(\gamma)>0$. Define
\begin{align}\label{Qhat}
\hat{\bQ}_i(\gamma)= (\hat{\bq}_{i,1}(\gamma), \ldots, \hat{\bq}_{i,k_i}(\gamma)),
\end{align}
{then} ${\cal M}(\bA_i)$ can be estimated by ${\cal M}(\hat{\bQ}_i(\gamma))$. When $\gamma$ is in regime $i$, similar to Theorem 1 in \cite{lam2011}, we can show that $\hat{\bQ}_i(\gamma)$ {provides a consistent estimator for the loading space under mild conditions}.

\begin{proposition}\label{prop:D}
{Assume Conditions 1--8 in Appendix A.1. If $p^{\delta_{\max}} n^{-1/2}=o(1)$, then as $n,p \to \infty$, with true $k_1$ and $k_2$, we have
\[
{\cal D}\{ \cM [\hat{{\bQ}}_1(\gamma)],\, {\cM}(\bQ_1) \}= O_p(p^{\delta_1} n^{-1/2})
\]
when $\gamma \leq \gamma_0$, and
\[
{\cal D} \{ \cM [\hat{\bQ}_2(\gamma)],\, {\cM}(\bQ_2) \}= O_p(p^{\delta_2} n^{-1/2})
\]
when $\gamma \geq \gamma_0$, where $\delta_{\max} = \max(\delta_1,\delta_2)$.}
\end{proposition}
{We remark that} when $\delta_i=0$, the estimator $\cM[\hat{\bQ}_i(\gamma)]$ converges to $\cM(\bQ_i)$ at {the} rate of $n^{-1/2}$, and {thus} the curse of dimensionality does not exist. When the factors in regime $i$ are weak, {however,} the convergence rate is slower and the noise process distorts the information on the latent factor; {see for example \cite{lam2011}}.

{We shall here also provide a discussion about the situation when $\gamma$ does not fall in regimes $i$. Without loss of generality, we illustrate by using the case when $\gamma \geq \gamma_0$. In this case,} $\bM_2(\gamma)$ is sandwiched by $\bA_2$, and $\hat{\bQ}_2(\gamma)$ is a reasonable estimate for $\cM(\bA_2)$. However, misclassification does occur for the estimation of regime 1. {In particular, since data points with $\lfloor \gamma n\rfloor < t \leq r_0$ from regime 2 are included in the calculation of ${\bM}_1(\gamma)$, it is no longer sandwiched by $\bA_1$. However, when $\gamma$ is sufficiently close $\gamma_0$, one can show that the misclassification effect becomes negligible on the estimation, and the estimated space using the sample version of $\bM_{1}(\gamma)$ will continue to be consistent; see the results in Section \ref{subsec:cpestimation}.}

\subsection{Estimating the Change Point Location}\label{subsec:cpestimation}
{In our estimation approach, we assume that the change point does not happen in the boundary area, namely there exists $0 < \eta_1 < \eta_2 < 1$ such that $\gamma_0 \in (\eta_1,\eta_2)$. Let $\gamma$ be a hypothesized change point location, then we can use it to split the data into two subsets, namely the one with $t \leq \lfloor \gamma n \rfloor$ and the other with $t > \lfloor \gamma n \rfloor$, and we define
\begin{equation}
G(\gamma)=\sum_{i=1}^2  g_i(\gamma),\quad g_i(\gamma)=\Big\|  {\bB_i(\eta_i)}' \, \bM_i(\gamma) \, \bB_i(\eta_i)\Big\|_2, \label{GG}
\end{equation}
where $\bB_i$ is a $p \times (p-k_i)$ matrix for which $(\bQ_i, \bB_i)$ forms a $p \times p$ orthonormal matrix with $\bQ_i' \bB_i=\mathbf{0}$ and $\bB_i' \bB_i= \bI_{p-k_i}$. In this case, ${\cM}(\bB_i)$ represents the orthogonal complement space of ${\cM}(\bQ_i)$ for $i=1,2$.} Note that although $\bB_i$ is not uniquely defined and subject to any orthogonal transformation, $g_i(\gamma)$ is invariant under such transformations. {If we} project the cross moment matrices $\{\bSigma_{y,i}(h, \gamma),\ h=\pm 1,\ldots, \pm h_0\}$ onto $\cM(\bB_i)$, then by (\ref{Mi}) we can see that $G(\gamma)$ measures the squared norm of the projections. If $\gamma=\gamma_0$ {is correctly specified}, then the data in the two regimes identified by $\gamma$ do belong to the correct regimes. Since $\bM_i(\gamma_0)=\bM_i$, $i=1,2$, by the definition of $\bM_i$ in (\ref{M}), we have
\[
G(\gamma_0)=\sum_{i=1}^2  \Big\|  {\bB_i}' \, \bM_i(\gamma_0) \, \bB_i\Big\|_2= \sum_{i=1}^2 \Big\| \sum_{h=1}^{ h_0}\left\{ {\bB_i}' \, \bA_i \left[ \bSigma_{x,i}(h,\gamma_0)\bA_i'\bA_i \bSigma_{x,i}(h,\gamma_0)\right]\bA_i' \, \bB_i\right\} \Big\|_2=0.
\]
If $\gamma \neq \gamma_0$, the data are not correctly separated and at least one of the subsets contains data from both regimes. {The} following proposition shows that under mild conditions $G(\gamma)>0$ for $\gamma \neq \gamma_0$.
\begin{proposition}
Under Conditions 1--8, $G(\gamma)>0$ if $\gamma \neq \gamma_0$.
\end{proposition}

Since $\gamma_0 \in (\eta_1, \eta_2)$, we can use data
corresponding to $t \leq \lfloor \eta_1n \rfloor $ and $t > \lfloor \eta_2 n \rfloor $ to get consistent
estimates for ${\cal M}(\bB_{1})$ and ${\cal M}(\bB_{2})$. Specifically, {for $i = 1,2$, we estimate $\bB_i$ by
\[
\hat{\bB}_i(\eta_i)=(\hat{\bq}_{i,k_i+1}(\eta_i), \ldots, \hat{\bq}_{i,p}(\eta_i)),
\]
where $\hat{\bq}_{i,k}(\eta_i)$ is the unit eigenvector of $\hat{\bM}_i(\eta_i)$ corresponding to the $k$-th largest eigenvalue. Then the sample version of the objective function $G(\gamma)$ is given by
\begin{equation}\label{g}
\hat{G}(\gamma)=\sum_{i=1}^2  \Big\|  \hat{\bB}_i(\eta_i)' \, \hat{\bM}_i(\gamma) \, \hat{\bB}_i(\eta_i) \Big\|_2.
\end{equation}
We propose to estimate the change point location $\gamma_0$ by
\begin{equation}\label{threshold}
\hat{\gamma} = \mathop{\mathrm{argmin}}_{\gamma \in \{0, \frac{1}{n}, \ldots, 1 \} \cap (\eta_1, \eta_2)}
\hat{G}(\gamma),
\end{equation}
whose asymptotic property is provided in Theorem \ref{thm:gammahat}.}

\begin{theorem}\label{thm:gammahat}
{Assume Conditions 1--8 in Appendix A.1. If $p^{\delta_{\max}}n^{-1/2}=o(1)$, then for any $\epsilon>0$, with true $k_1$ and $k_2$ we have}
\begin{equation*}
P(\hat{\gamma}<\gamma_0-\epsilon)\leq \frac{Cp^{\delta_1}}{ \epsilon n^{1/2}}, \quad
P(\hat{\gamma}>\gamma_0+\epsilon)\leq \frac{Cp^{\delta_2}}{ \epsilon n^{1/2}},
\end{equation*}
as $n,p \to \infty$.
\end{theorem}

{By Theorem \ref{thm:gammahat}, the proposed estimator $\hat \gamma$ in (\ref{threshold}) for the change point location is consistent under mild conditions. It also reveals that the estimation performance can depend critically on the strength of factors in both regimes. In particular, if} the factors are strong in both regimes ($\delta_1=\delta_2=0$), {then} the estimation is immune to the curse of dimensionality. {On the other hand}, if factors are weak in one regime, {then} the {resulting} estimator {can become less efficient as $p$ increases.} When factors have different levels of strengths before and after {the} break, the probability that the $\hat{\gamma}_0$ falls in the weaker regime is larger but
the estimation precision in the stronger regime is better. {As a result,} the overall rate of convergence of $\hat{\gamma}$ depends on the strength of the
weaker regime.

{By plugging $\hat \gamma$ into the estimation procedure described in Section \ref{subsec:estimation}, we obtain the final estimators for ${\cal M}(\bA_i)$. To be more specific, let $\hat r = \lfloor n \hat \gamma \rfloor$ and}
\[
\hat{\bQ}_i(\hat{r})=(\hat{\bq}_{i,1}(\hat{r}), \ldots, \hat{\bq}_{i,k_i}(\hat{r})),
\]
where $\hat{\bq}_{i,k}(\hat{r})$ is the unit eigenvector of
$\hat{\bM}_i(\hat{r})$ corresponding to its $k$-th largest eigenvalue.
Theorem 2 provides the asymptotic property of the estimated loading spaces when the estimated break date is used.

\begin{theorem}\label{thm:Dwgammahat}
{Assume Conditions 1--8 in Appendix A.1. If $p^{\delta_{\max}}n^{-1/2}=o(1)$, then as $n, p\to \infty$, with true $k_1$ and $k_2$, we have
\begin{equation*}
{\cal D}\{ {\cal M}[ \hat{\bQ}_i(\hat{\gamma})] ,\,{\cal M}(\bQ_i)\} = O_p(p^{\delta_i}n^{-1/2})
\end{equation*}
for $i = 1,2$.}
\end{theorem}

{By Theorem \ref{thm:Dwgammahat}, the convergence rate of the associated loading space estimators is the same as that in Proposition \ref{prop:D} when the true change point location is known. Compared with the results in \cite{liu} which used a threshold variable to split the data}, the `helping effect' disappears and asymptotically there is no interaction between regimes. {In practice, the number of factors is typically unknown but can be estimated through a similar eigenvalue ratio estimator as that used in \cite{lam2011}, namely
\begin{equation}\label{est_num}
\hat{k}_i = \mathop{\mathrm{argmin}}_{1 \leq k \leq p/2}
\frac{{\hat{\lambda}_{i,k+1}(\eta_i)}}{{\hat{\lambda}_{i,k}(\eta_i)}},
\end{equation}
where $\hat{\lambda}_{i,k}(\eta_i)$ is the $k$-th largest eigenvalue of $\hat{\bM}_i(\eta_i)$, $i=1,2$.}

\section{Determining the Existence of Change Point}\label{sec:cptest}
\subsection{A Self-Normalized Change-Point Test}\label{subsec:cptest}
Although the change-point factor model (\ref{model}) is able to capture potential structural breaks in the loading space, it can be unnecessarily complicated when there is actually no change point. We shall here consider the problem of determining the existence of change points in the high-dimensional factor model (\ref{model}). The problem has been studied by \citet{Chen:Dolado:Gonzalo:2014} using linear regression of estimated factors, and by \citet{Ma:Su:2018} using fused Lasso penalization on block estimators; see also reference therein. However, the aforementioned results both require the noise vector to have a sparse cross-sectional dependence structure; see for example Assumption 4 in \citet{Chen:Dolado:Gonzalo:2014} and Assumption A3 in \citet{Ma:Su:2018}. In addition, the fused Lasso approach of \citet{Ma:Su:2018} does not provide a $p$-value which can be an informative measure for quantifying the amount of statistical evidence in favor of the change-point model. To handle noise processes with possibly non-sparse cross-sectional dependence structures, the main challenge is to deal with the unknown scale parameter that associates with the strength of cross-sectional dependence. For this, we propose to adopt the idea of self-normalization \citep{Lobato:2001,Shao:2010}, which is capable of adaptively handling unknown scale parameters caused by different dependence strengths; see for example \citet{Shao:2011:CPLRD}, \citet{Bai:Taqqu:Zhang:2016} and \citet{Taqqu:Zhang:2019}.

Assume that the noise process $(\bve_t)$ has a constant covariance structure $\bSigma_t \equiv \bSigma$, then for any vector $\bb \in {\cal M}(\bB_1)$, by (\ref{model}) and Condition 3 in Appendix A.1 we have
\begin{equation*}
\mathrm{var}(\bb' \by_t) = \bb' \bSigma \bb,\quad t \leq r_0,
\end{equation*}
and
\begin{equation*}
\mathrm{var}(\bb' \by_t) = \mathrm{var}(\bb' \bA_2\bx_{t,2} + \bb' \bve_t) = \bb' \bA_2 \mathrm{cov}(\bx_{t,2}) \bA_2'\bb + \bb' \bSigma \bb,\quad t > r_0.
\end{equation*}
Therefore, one can determine the existence of change points by testing for structural breaks in the variance of the transformed sequence $(\bb' \by_t)$, namely
\begin{equation}\label{eqn:H_0}
H_0:\ \mathrm{var}(\bb' \by_1) = \cdots = \mathrm{var}(\bb' \by_n)
\end{equation}
versus the alternative that there exists a time point $r_0$ such that $\mathrm{var}(\bb' \by_{r_0}) \neq \mathrm{var}(\bb' \by_{r_0+1})$. Compared with traditional change-point testing problems, the key difference here is that the scale of the random variable sequence $(\bb' \by_t)$ can grow to infinity at an unknown rate depending on the strength of cross-sectional dependence. This is particularly due to the fact that the dimension of the covariance matrix $\bSigma$ is $p \times p$, where $p$ can grow to infinity with the sample size $n$. Under the idiosyncratic error assumption of \citet{Chen:Dolado:Gonzalo:2014} and \citet{Ma:Su:2018}, the covariance matrix $\bSigma$ has a sparse structure by which one can show that $\mathrm{var}(\bb' \by_t) = O(\|\bb\|_2^2)$. However, for situations with non-sparse cross-sectional dependence structures, $\mathrm{var}(\bb' \by_t)$ can grow to infinity even when $\|\bb\|_2$ is bounded, and the rate can depend on the strength of the underlying cross-sectional dependence, which is typically unknown in practice. As a result, this can pose a challenge when one seeks an appropriate normalization for the test statistic. For this, we consider the following self-normalization approach.

Let
\begin{equation}\label{eqn:nu}
\hat \nu_{i,j} = (j-i)^{-1}\left(\sum_{t=i}^j  \bb'  \by_t \by_t' \bb- \sum_{t=i}^j   \bb' \by_t \sum_{t=i}^j \by_t'\bb \right) ,
\end{equation}
we consider the test statistic
\begin{equation}\label{eqn:Tn}
T_n = \max_{n\eta_1 < r < n\eta_2} {\{r(n-r)(\hat \nu_{1,r} - \hat \nu_{r+1,n})\}^2 \over n^2V_r},
\end{equation}
where
\begin{equation*}
V_r = {1 \over n} \left[\sum_{i=1}^r \left\{{i(r-i)(\hat \nu_{1,i} - \hat \nu_{i+1,r}) \over r}\right\}^2 + \sum_{i=r+1}^n \left\{{(i-r-1)(n-i+1)(\hat \nu_{r+1,i-1} - \hat \nu_{i,n}) \over n-r}\right\}^2\right]
\end{equation*}
is the self-normalizer that pivotalizes the asymptotic distribution of the test statistic (\ref{eqn:Tn}) to make it free of any nuisance scale parameter.

\begin{theorem}\label{thm:cptest}
Assume Condition 9 in Appendix A.1. Then under the null hypothesis of no change point, we have as $n, p \to \infty$,
\begin{equation*}
T_n \to_D \sup_{s \in (\eta_1,\eta_2)} {\{\mathbb B(s) - s\mathbb B(1)\}^2 \over W(\mathbb B,s)},
\end{equation*}
where
\begin{equation*}
W(\mathbb B,s) = \int_0^s \left\{\mathbb B(u) - {u \over s}\mathbb B(s)\right\}^2du + \int_s^1 \left[\mathbb B(1) - \mathbb B(u) - {1-u \over 1-s} \{\mathbb B(1) - \mathbb B(s)\}\right]^2du.
\end{equation*}
\end{theorem}

\subsection{A Data-Driven Choice}\label{subsec:teststat}
{Although the hypothesis test in (\ref{eqn:H_0}) is able to detect the change point with any $\bb \in \cM(\bB_1)$, we shall here consider a data-driven choice that aims to optimize the power performance. Intuitively, $\bb$ should be chosen to be far away from the linear space $\cM(\bB_2)$, which can} be achieved by
\[
\max_{\bb \in \cM(\bB_1)} \| \bb' \bB_2 \|_2.
\]
{As a result, we propose to choose} $\bb$ by
\[
\max_{\bu \in \mathbb{R}^{k_1}, \|\bu\|_2=1 } \|\bu' \bB_1' \bB_2\|_2, \mbox{ where } \bb=\bB_1 \bu.
\]
When $\bu$ is the unit right singular vector of $\bB_2'\bB_1$ {that corresponds} to the largest singular value, $\bb$ {will be} in the column space of $\bB_1$ and the distance of $\bb$ and ${\cM}(\bB_2)$ {will then be} maximized. Let $\hat{\bu}(\eta_1,\eta_2)$ be the unit right singular vector of $\hat{\bB}_2'(\eta_2)\hat{\bB}_1(\eta_1)$ {that corresponds to the} largest singular value, {the associated test statistic with the data-driven choice can then be obtained by replacing $\bb \by_t$ with $\bB_1\hat{\bu}(\eta_1,\eta_2)$ in (\ref{eqn:nu}) and (\ref{eqn:Tn}). When the degrees of factor strength are different in the two regimes, the data from the one where factors are stronger will provide more information. We can further incorporate this into the data-driven choice of $\bb$. To be more specific, if $\|{\hat{\bM}}_2(\eta_2)\|_2>\|{\hat{\bM}}_1(\eta_1)\|_2$, then factors after the potential change point are likely to be stronger, in which case we can choose $\bb \in {\cM}({\bB_2})$ that maximizes the distance from ${\cM}(\bB_1)$ by letting $\bb=\bB_2 \bv$, where $\bv$ is the right singular vector of $\bB_1'\bB_2$ corresponding to the largest singular value.}

\section{Simulation}\label{sec:numericalexperiments}
In this section, we present {a Monte Carlo simulation study to examine the finite sample performance of the proposed inference procedure. For this, we generate entries of the loading matrix $\bA_i$ as independent sample from the uniform distribution on $[-p^{-\delta_i/2}, p^{-\delta_i/2}]$. As a result, the factor strength of $\bA_i$ is characterized by $\delta_i$. The associated noise process is generated as a Gaussian process whose covariance matrix has 1 on the diagonal and $\rho_e$ for all off-diagonal entries. For simplicity, we use $h_0=1$ throughout our numerical examples. For each setting, we generate 1000 realizations, and the results are summarized in the following subsections.}

\subsection{The Case with No Change Point}\label{size}
{We first consider the case with no change point and examine the empirical size performance of the self-normalized change-point test proposed in Section \ref{sec:cptest}. For this, we consider both the setting with strong factors ($\delta = 0$) and the setting with weak factors ($\delta = 0.25$). Three common factors drive the time series, and the factor process is set to be three independent AR(1) processes with $N(0,4)$ noises process and AR coefficients 0.9, -0.7, and 0.8. $\rho_e=0.5$ for the noise process. $\eta_1=0.1$ and $\eta_2=0.9$. The results are summarized in Table \ref{size_ex1}, from which we can see that the empirical sizes of the proposed test are reasonably close to their nominal levels $\alpha = 10\%, 5\%, 1\%$ when factors are strong.} When $\delta=0.25$, our tests are slightly oversized, and this is because we only utilize information obtained before time $\lfloor  \eta_1n \rfloor$ and after time $\lfloor \eta_2 n \rfloor$ to estimate $\bb$ in (\ref{eqn:nu}).


\begin{table}[htbp!]
\begin{center}
\footnotesize
\caption{Empirical sizes of tests with different levels of factor strength in Section \ref{size}}
\begin{tabular}{cc |cc |cc | cc }
\hline \hline
		&&\multicolumn{2}{c|}{$\alpha=10\%$}		&\multicolumn{2}{c|}{$\alpha=5\%$}			& \multicolumn{2}{c}{$\alpha=1\%$}	\\ \hline
\multicolumn{2}{c|}{$\delta$} 	&0		&0.25&0	&0.25 &0	&0.25 \\ \hline

n=400&p=20		&0.112	&0.138	&0.064	&0.072	&0.014	&0.012\\
n=400&p=40		&0.127	&0.149	&0.069	&0.083	&0.013	&0.020\\
n=400&p=100 		&0.123	&0.112	&0.067	&0.059	&0.018	&0.016\\
n=1000&p=20		&0.116	&0.122	&0.058	&0.057	&0.009	&0.011\\
n=1000&p=40		&0.103	&0.130	&0.058	&0.072	&0.012	&0.012\\
n=1000&p=100		&0.115	&0.137	&0.068	&0.071	&0.014	&0.019\\
\hline \hline
\end{tabular}
\label{size_ex1}
\end{center}
\end{table}

\subsection{The Change-Point Case}\label{power}
{We now consider the change-point case and examine if the self-normalized change-point test proposed in Section \ref{sec:cptest} can successfully detect the existence of the change point. For this, we follow the data generating mechanism described in Section \ref{size} for generating the factor and noise processes, and set $k_1 = k_2 = 3$. Let the change point location $\gamma_0 = 0.5$, and we consider four different scenarios on the factor strength, namely SS ($\delta_1 = \delta_2 = 0$) in which strong factors are used before and after the change point, SW ($\delta_1 = 0$ and $\delta_2 = 0.25$) in which strong factors are used before the change point and weak factors after, WS ($\delta_1 = 0.25$ and $\delta_2 = 0$) in which weak factors are used before the change point and strong factors after, and WW ($\delta_1 = \delta_2 = 0.25$) in which weak factors are used before and after the change point. The results are summarized in Table \ref{power_ex2}, from which we can see that the proposed test performs reasonably well as it successfully identifies the change point with high probabilities. In addition, the performance seems to improve in general when the sample increases.}

\begin{table}
\footnotesize
\begin{center}
\caption{Empirical powers of tests under different settings in Section \ref{power}}
\begin{tabular}{cc |cccc |cccc | cccc }\hline \hline
&		&\multicolumn{4}{c|}{$\alpha=10\%$}		&\multicolumn{4}{c|}{$\alpha=5\%$}			& \multicolumn{4}{c}{$\alpha=1\%$}	\\ \hline

\multicolumn{2}{c|}{Setting} 	&SS	&SW	&WS	&{WW} &SS	&SW	&WS	&{WW} &SS	&SW	&WS	& {WW} \\ \hline
n=400	&p=20	&0.975	&0.973	&0.966	&0.969	&0.941	&0.929	&0.932	&0.943	&0.820	&0.787	&0.815	&0.810\\
n=400	&p=40	&0.979	&0.970	&0.955	&0.969	&0.944	&0.936	&0.915	&0.934	&0.820	&0.804	&0.783	&0.799\\
n=400	&p=100	&0.960	&0.968	&0.975	&0.977	&0.931	&0.924	&0.949	&0.940	&0.805	&0.806	&0.811	&0.821\\
n=1000	&p=20	&0.996	&0.996	&0.998	&0.996	&0.990	&0.989	&0.992	&0.992	&0.956	&0.949	&0.958	&0.959\\
n=1000	&p=40	&0.998	&0.996	&0.994	&0.997	&0.989	&0.987	&0.985	&0.987	&0.935	&0.956	&0.946	&0.950\\
n=1000	&p=100	&0.999	&0.990	&0.997	&0.999	&0.996	&0.995	&0.992	&0.992	&0.960	&0.962	&0.960	&0.962\\
\hline \hline
\label{power_ex2}
\end{tabular}
\end{center}
\end{table}

\begin{figure}
\centering
\includegraphics[width=6.5in]{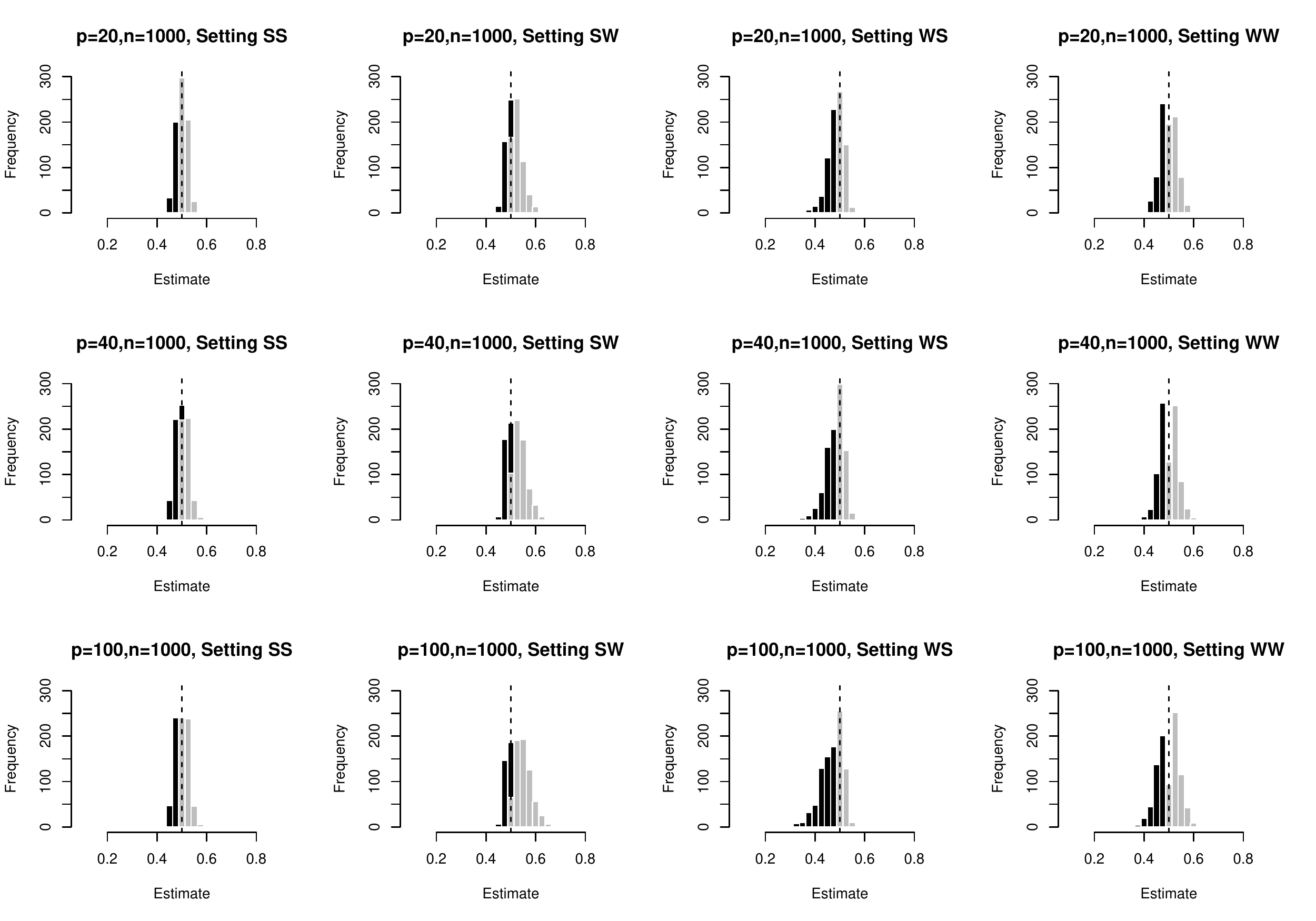}
\caption{Histograms of estimated threshold value under different settings when $n=1000$ and $k_1$ and $k_2$ are known. The dashed line shows the true threshold value $\gamma_0=0.5$, black bars show the frequencies of underestimation, and grey bars show the frequencies of overestimation. }
\label{simu:thres_est}
\end{figure}

{After the change point existence has been identified by the test, we shall apply the methods proposed in Section \ref{sec:estimation} to estimate the change point location and the loading spaces in two regimes. Figure \ref{simu:thres_est} provides the histograms of our change point location estimator $\hat \gamma$ under different settings, from which we can observe the followings. First, when the factor strength is weak in at least one regime, before or after the change point, the estimation efficiency in that weak regime can suffer from the increase in dimension. In contrast, the estimation efficiency in the strong regime does not seem to be affected by the curse of dimensionality. This is in line with the results in Theorem \ref{thm:gammahat}; see also the discussions thereafter. Second, it can be seen from the top panels in Figure \ref{simu:thres_est} that, when the factor strengths before and after the change point are different, namely settings SW and WS, the estimation bias, though asymptotically negligible, is more likely to be toward the regime with weaker factors. In particular, when the factor strength after the change point is weaker as in the SW setting, then it is more likely to overestimate $\gamma_0$. On the other hand, if the factor strength before the change point is weaker as in the WS setting, then it is more likely to underestimate $\gamma_0$. We also provide in Table \ref{tab:gammahat} a summary of the estimation error $|\hat{\gamma}-\gamma_0|$. The estimation error for the loading spaces are summarized in Table \ref{tab:D}, from which we can see that the estimation procedure proposed in Section \ref{subsec:estimation} performs reasonably well under all the considered settings.}

\begin{table}[htbp!]
\caption{Average estimation error $|\hat{\gamma}-\gamma_0|$ in Section \ref{power}}
\centering
\begin{tabular}{l | ccc |ccc}
\hline \hline
$n$	    &\multicolumn{3}{c|}{$n=400$} &\multicolumn{3}{c}{$n=1000$}  \\ \hline
$p$	         &$20$  &$40$	&$100$     &$20$  &$40$  &$100$   \\ \hline
SS    	&0.035	&0.039	&0.040	&0.015	&0.018	&0.018			 \\ \hline
SW   	&0.051	&0.066	&0.083  	&0.023	&0.029	&0.039			 \\ \hline
WS   	&0.054	&0.060	&0.081 	&0.023	&0.027	&0.038		 \\ \hline
WW    	&0.053	&0.060	&0.071	&0.024	&0.028	&0.034		  \\
\hline \hline
\end{tabular}\label{tab:gammahat}
\end{table}

\begin{table}[htbp!]
\caption{Average estimation error {$\cD\{\cM [\hat{\bQ}_i(\hat{r})],\, \cM(\bQ_i)\}$} in Section \ref{power}}
\centering
\begin{tabular}{l| l| ccc |ccc}
\hline \hline
&$n$	    &\multicolumn{3}{c|}{$n=400$} &\multicolumn{3}{c}{$n=1000$}  \\ \hline
&$p$	         &$20$  &$40$	&$100$     &$20$  &$40$  &$100$   \\ \hline
Setting SS  & $\delta_1=0$  	&0.059	&0.055	&0.053	&0.033	&0.032	&0.031 \\
	      &$\delta_2=0$   		&0.059	&0.057	&0.055	&0.033	&0.032	&0.031 \\ \hline
Setting SW  & $\delta_1=0$  	&0.058	&0.056	&0.053	&0.033	&0.032	&0.031 \\
	      &$\delta_2=0.25$ 	&0.095	&0.099	&0.113	&0.049	&0.053	&0.058 \\ \hline
Setting WS  & $\delta_1=0.25$ 	&0.093	&0.094	&0.110	&0.048	&0.052	&0.058 \\
	      &$\delta_2=0$   		&0.060	&0.056	&0.053	&0.034	&0.032	&0.032 \\ \hline
Setting WW &$\delta_1=0.25$   &0.089	&0.095	&0.103	&0.049	&0.052	&0.057\\
	     &$\delta_2=0.25$    	&0.093 	&0.094	&0.105	&0.050	&0.052	&0.057 \\ \hline \hline	
\end{tabular}\label{tab:D}
\end{table}

\subsection{A Comparison with Existing Results}\label{comparison}

In this subsection, we will compare the performance of our change point detection procedure with \citet{Chen:Dolado:Gonzalo:2014} and \citet{han2015} who both focus on single change point detection in factor models but do not allow strong cross-sectional dependence in noise process. Set $\eta_1=0.2$ and $\eta_2=0.8$. Data are simulated with processes described in Section \ref{size} and in Section \ref{power} with $\rho_e=0.95$ to compare the sizes and powers of different tests, respectively. From Table \ref{comp_ex1_size} and Table \ref{comp_ex1_power}, we can see that our test controls the sizes better than Wald test proposed by \cite{Chen:Dolado:Gonzalo:2014} and is more powerful than LM and Wald tests by \cite{han2015} when the strong cross-sectional exists in noise process.

\begin{landscape}
\begin{table}[htbp!]
\footnotesize
\caption{Size comparison among different tests with $\alpha=5\%$ for Example 1 in Section \ref{comparison}}
\begin{center}
\begin{tabular}{l| l| cc |cc| cc| cc}
\hline \hline
\multicolumn{2}{c|}{$\delta$}    &\multicolumn{2}{c|}{Our method}  	&\multicolumn{2}{c|}{Wald-cdg}   &\multicolumn{2}{c|}{LM-hi} &\multicolumn{2}{c}{Wald-hi}   \\ \hline
$n$	&$p$	&0 &0.25&0		&0.25 &0		&0.25 &0		&0.25  \\ \hline
$100$& $20$ 	&0.086	&0.100	&0.238	&0.203	 &0.016	&0.017	&0.118	&0.130\\
 $100$& $40$  	&0.110	&0.148	&0.267	&0.203	&0.004	&0.010 	&0.161	&0.136\\
$100$& $100$ 	&0.119	&0.204	&0.235	&0.210	 &0.006	&0.008	&0.161	&0.141\\ \hline
$400$& $20$ 	&0.074	&0.071		&0.072	 &0.092	&0.028	&0.023	&0.045	&0.042\\
 $400$& $40$  	&0.086	&0.073		&0.072	&0.092	&0.029	&0.031 	&0.061	&0.045\\
$400$& $100$  &0.099	&0.077		&0.084	&0.093	&0.023	&0.026	&0.050	&0.050\\ \hline \hline	
\end{tabular}\label{comp_ex1_size}
\end{center}

\caption{Power comparison among different tests with $\alpha=5\%$ for Example 1 in Section \ref{comparison}}
\begin{center}
\begin{tabular}{l| l| cccc |cccc| cccc| cccc}
\hline \hline
\multicolumn{2}{c|}{$\delta_1,\delta_2$}    &\multicolumn{4}{c|}{Our method}  	&\multicolumn{4}{c|}{Wald-cdg}   &\multicolumn{4}{c|}{LM-hi} &\multicolumn{4}{c}{Wald-hi}   \\ \hline
$n$	&$p$	&0,0	&0,0.25	&0.25,0	&0.25,0.25&0,0	&0,0.25	&0.25,0	&0.25,0.25 &0,0	&0,0.25	&0.25,0	&0.25,0.25 &0,0	&0,0.25	&0.25,0	&0.25,0.25  \\ \hline
$100$& $20$ 	&0.721	&0.692	&0.671	&0.676	&0.960	&0.948	&0.950	&0.928	&0.036	&0.057	&0.024	&0.028	&0.109	&0.066	&0.164	&0.138	 \\
 $100$& $40$  	&0.722	&0.718	&0.700	&0.662	&0.999	&0.998	&0.999	&0.997	 &0.007	&0.006	&0.008	&0.003	&0.212	&0.168	&0.226	&0.187\\
$100$& $100$ 	&0.715	&0.700	&0.710	&0.681	&0.859	&0.809	&0.809	&0.821	 &0.000	&0.000	&0.000	&0.000	&0.250	&0.210	&0.260	&0.264\\ \hline
$400$& $20$ 	&0.930	&0.929	&0.921	&0.925	&1.000	&0.996	&0.992 	&0.995	&1.000	&1.000	&1.000	&0.995 &1.000	&1.000	&0.999	&0.996\\
 $400$& $40$  	&0.951	&0.919	&0.924	&0.913	&1.000	&1.000	&1.000	&1.000	&0.995	&0.360	&0.382	&0.319	&0.396	&0.400	&0.393	&0.447\\
$400$& $100$  &0.939	&0.933	&0.912	&0.928	&1.000	&1.000	&1.000	&1.000	&0.323	&0.264	&0.318	&0.310	&0.396	&0.349	&0.392	&0.411\\ \hline \hline	
\end{tabular}\label{comp_ex1_power}
\end{center}
Note: 'cdg' denotes the test proposed by \cite{Chen:Dolado:Gonzalo:2014}, 'hi' denotes the test proposed by \cite{han2015}, and 'LM' denotes Lagrange Multiplier test. Wald test in \cite{Chen:Dolado:Gonzalo:2014} is more powerful than their LM test. \cite{han2015} developed various test statistics which have similar performance. In Table 5 and Table 6, we compare our test with Wald test by \cite{Chen:Dolado:Gonzalo:2014}, LM and Wald tests by \cite{han2015} with supremum test statistics obtained by Bartlett kernel.
\end{table}

\end{landscape}

\section{Real data analysis}\label{sec:realdata}
We applied our method to the Stock-Watson data \citep{stock1998, stock2005}, containing 132 U.S. monthly economic indicators from March 1960 to December 2003, with $n=526$ and $p=132$. The data include real output and income, employment, real retail, manufacturing and trade sales, consumption, interest rates, price index and other economic indicators. \cite{stock2005} provided more detailed information about this data set and transformations needed for stationarity before analysis.

Set $h_0=1$, $\eta_1=0.1$, $\eta_2=0.9$, and level of significance $\alpha=5\%$. We applied the wild binary segmentation method \citep{Fryzlewicz:2014} to check if there are multiple change points. The result shows that during this period there is only one change point. 

\begin{figure}[h]\label{fig:eigenratio}
\centering
\includegraphics[width=5in]{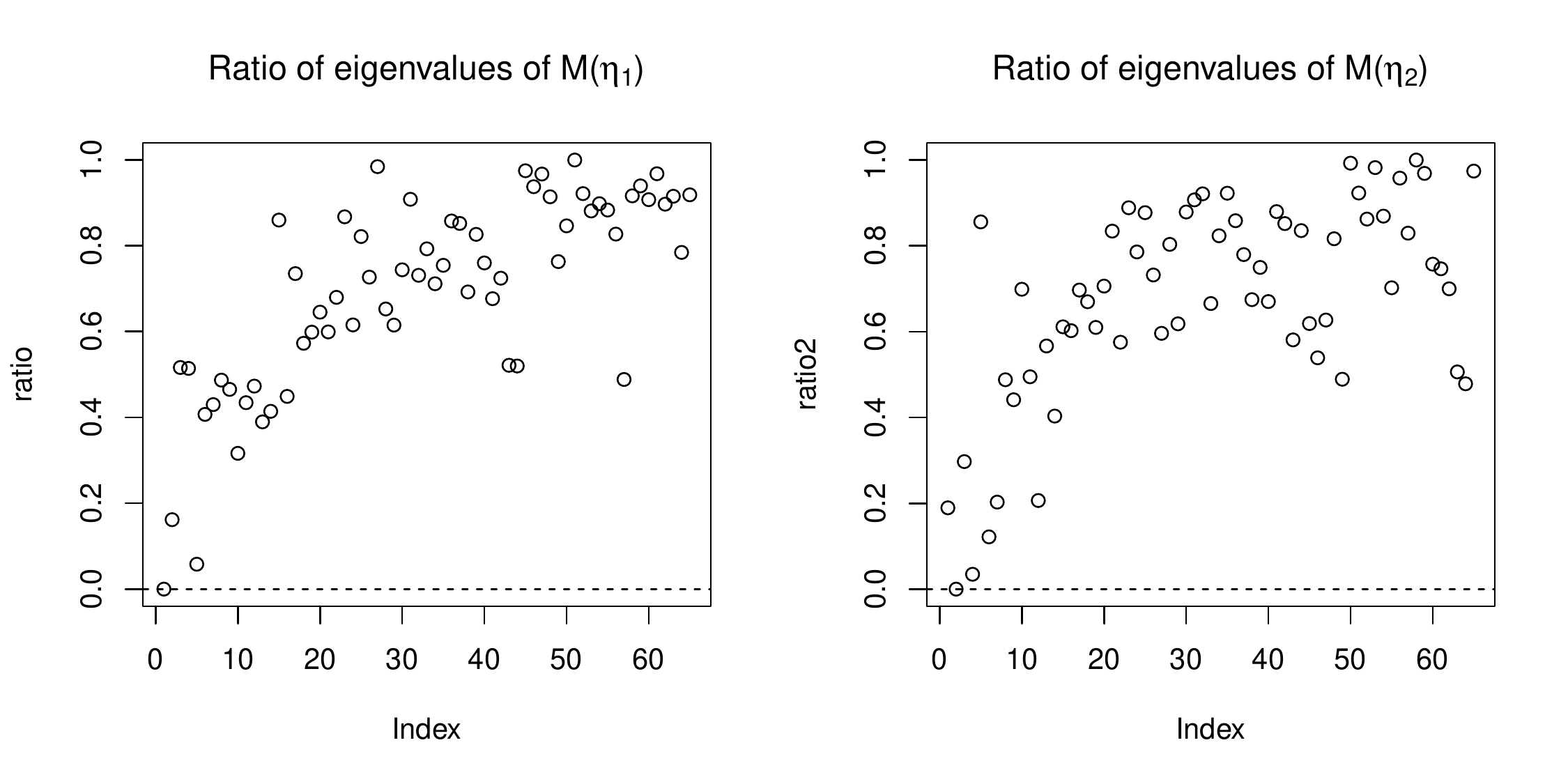}
\end{figure}

Left and right panels in Figure \ref{fig:eigenratio} demonstrate the ratio of eigenvalues of $\hat{\bM}(\eta_1)$ and $\hat{\bM}(\eta_2)$, respectively, and it reaches its minimum values at 1 and 2, which implies that $\hat{k}_1=1$ and $\hat{k}_2=2$. Using methods described in Section \ref{subsec:cpestimation}, we have $\hat{\gamma}=0.635$. 

 It means that the dynamics of economic indicators experienced once a permanent structural change around November 1988 possibly due to saving and loan crisis \citep{curry2000}. Before the change, there was one common factor, while after the change, there was two common factors driving 132 economic indicators.

 Extending the method described in \cite{lam2011}, we can estimate the residuals of factor models with
 \begin{align*}
 \hat{\bve}_t=\left\{
\begin{array}{cc}
 \hat{\bQ}_1\hat{\bQ}_1' \by_t	&t\leq \hat{\gamma}n, \\
  \hat{\bQ}_2\hat{\bQ}_2' \by_t	&t> \hat{\gamma}n.
 \end{array}
\right.
 \end{align*}
We compare our method with Wald test by \cite{Chen:Dolado:Gonzalo:2014}, LM and Wald tests by \cite{han2015} through white noise test \citep{chang2017} and residual sum of squares (RSS). All the three tests reject the null and agree that there is a structural break in data. With the estimated break dates, we fix the number of factors before and after break at 1 and 2 for comparison and calculate the residual series. Table \ref{table:comp} reports the $p$-values associated with white noise test for the residual series and RSS. Our model is validated by the white noise test with $p$-value 0.07. It confirms that after extracting the common factors, there is no serial dependence left. The residuals series from Wald test by \cite{Chen:Dolado:Gonzalo:2014} and Wald test by \cite{han2015} fail the white noise test. Although $p$-value for LM test by \cite{han2015} is greater than 0.05, \cite{han2015} yields a larger RSS. In sum, our method can successfully capture all the serial dynamics of the data, and lower the residual sum of squares comparing to others.

\begin{table}[h]
\begin{center}
\caption{Comparison among different models for real data analysis}
\begin{tabular}{l| cc}\hline \hline
Method		&$p$-value for white noise test	&RSS\\ \hline
Ours				&0.075	&308238\\
Wald-cdg		&0.000	&160476	\\
LM-hi		&0.070	&1482874	\\
Wald-hi		&0.030 &1156676 \\ \hline \hline	
\end{tabular}\label{table:comp}
\end{center}
\end{table}

\section{Conclusion}\label{sec:conclusion}
{Factor models have been frequently used in the study of high-dimensional time series. Their associated change-point analyses were mostly studied under the framework of \citet{bai2002} and \citet{forni2004}. However, such type of factor models suffer from the identifiability issue in the finite dimensional case and cannot be used to explain the phenomenon of strong cross-sectional dependence in noise process. To alleviate the aforementioned problems, \citet{lam2011} proposed an alternative framework for factor analysis of high-dimensional time series. In the current paper, we develop statistical methods and their associated theory for change-point analysis of the recently proposed factor model of \citet{lam2011}. In the presence of a change point, we propose consistent statistical estimators for the change point location and factor loading spaces before and after the change point, and provide their explicit convergence rates. In particular, our results reveal that the resulting estimators can have very different asymptotic behaviors in response to the phenomenon of high dimensionality depending on the factor strength. To be more specific, for processes with strong factor strength, the convergence rate will not be affected by the dimension, while the curse of dimensionality can be observed for processes with weak factor strength; see the discussion in Section \ref{sec:estimation} and our numerical results in Section \ref{sec:numericalexperiments}. We also propose a self-normalized test for determining the change-point existence, which, due to the self-normalization, can adaptively handle the case with strong and weak cross-sectional dependence. It can be seen from our numerical results in Section \ref{sec:numericalexperiments} that the proposed change-point test performs reasonably well in terms of both the size and power.}

\section*{Appendix A.1. Regularity Conditions}
Define
\[
N(c_1,c_2)=\lfloor c_1n \rfloor -\lfloor c_2 n \rfloor,
\]
\[
\bx_t=\sum_{i=1}^2 \bx_{t,i}I_{t,i},
\]
and three intervals
\[
I_1(h)=[0,\gamma_0-h/n], \quad I_2(h)=(\gamma_0-h/n,\gamma_0], \quad I_3(h)=(\gamma_0,1].
\]
For any $0 \leq c_1< c_2 \leq 1$ and $c_1$ and $c_2$ are from the same one of the three intervals, $I_1$, $I_2$ or $I_3$, let
\[
{\bGamma}_x(h, c_1,c_2)=\frac{\sum_{t=1}^{n} {\rE} (\bx_{t} \bx_{t+h}' I_{\{ \lfloor c_1n \rfloor< t \leq \lfloor c_2 n \rfloor \} })}{ N(c_1,c_2)}.
\]

\noindent
The following regularity conditions are needed for theoretical properties.

\noindent \textbf{Condition 1.} Let ${\cal F}_{\ell}^{j}$ be the $\sigma$-field generated by $\{(\bx_{t,1}, \bx_{t,2}): \ell \leq t \leq j\}$
The latent process $\{\bx_{t,1}, \bx_{t,2}\}$ is
$\alpha$-mixing with mixing coefficients satisfying
\[
\sum_{t=1}^{\infty} \alpha(t)^{1-2/\zeta}<\infty,
\]
for $i=1,2$ and some $\zeta>2$, where
$\alpha(t)=\sup_{j} \sup_{A \in \calF_{-\infty}^j, B \in \calF_{j+t}^{\infty}} |P(A \cap B) -P(A)P(B)|$.

\noindent \textbf{Condition 2.} For any $i=1,2$, $j=1,\ldots, k_i$, and $t=1, \ldots, n$, $E(|x_{t,i,j}|^{4\zeta})< \sigma_x^{4\zeta}$, where $x_{t,i,j}$ is the $j$-th element of $\bx_{t,i}$, $\sigma_x>0$ is a constant, and $\zeta$ is given in Condition 1. 

\noindent \textbf{Condition 3.} $\{\bve_t\}$ is a white noise process. $\bve_{t}$ and $\{\bx_{t,1}, \bx_{t,2}\}$ are uncorrelated  given ${\cal F}_{-\infty}^{t-1}$. Each element of $\bSigma$ remains bounded by a positive constant $\sigma_\varepsilon^2$ as $p$ increases to infinity.

\noindent \textbf{Condition 4.} For $i=1,2$, there exists a constant $\delta_i \in [0,1]$ such that $\|\bA_i\|_2^2 \asymp \|\bA_i\|^2_{\min}  \asymp p^{1-\delta_i} $, as $p$ goes to infinity.

\noindent \textbf{Condition 5.} $\gamma_0 \in (\eta_1,\eta_2)$. For any $\gamma \in [\eta_1,\eta_2]$, there exists an integer $h_i \in [1, h_0]$ such that $\bSigma_{x,i}(h_i, \gamma)$ and $\bSigma_{x,i}(h_i, \gamma)$ are full rank, and $\|\bSigma_{x,i}(h_i,\gamma)\|_{\min}$ is uniformly bounded above 0, for $i=1,2$.

\noindent \textbf{Condition 6.} $\bM_i(\gamma)$ admits $k_i$ distinct positive eigenvalues, for $\gamma \in [\eta_1,\eta_2]$, $i=1,2$.

\noindent \textbf{Condition 7.} There exists a positive constant $d$ such that ${\cal D}({\cal M}(\bQ_1), {\cal M}(\bQ_2))> d$ as $n$ and $p$ go to infinity.

\noindent \textbf{Condition 8.} For any $\gamma \in (\eta_1,\gamma_0)$, there exists an integer $h_1^*\in [1, h_0]$ such that $\bGamma_x(h_1^*,\gamma, \gamma_0-h_1^*/n)$ is full rank. For any $\gamma \in (\gamma_0, \eta_2)$, there exists an integer $h_2^* \in[1,h_0]$ such that $\bGamma_x(h_2^*, \gamma_0, \gamma)$ is full rank. The minimum singular values of these two matrices mentioned are uniformly bounded above $u_0>0$.


Let
\begin{equation*}
\hat \nu_{i,j}^\star = (j-i)^{-1} \bb' \left(\sum_{t=i}^j \bve_t \bve_t' - \sum_{t=i}^j \bve_t \sum_{t=i}^j \bve_t'\right) \bb,
\end{equation*}
and we use $\rightsquigarrow$ to denote the weak convergence of processes \citep{vanderVaart:Wellner:1996}.

\noindent \textbf{Condition 9.} There exists a deterministic sequence $(\sigma_{n,p})$ such that
    \begin{equation*}
    \{\sigma_{n,p}^{-1}(\lfloor nt \rfloor - \lfloor ns \rfloor \vee 1 + 1) (\hat \nu_{\lfloor ns \rfloor \vee 1,\lfloor nt \rfloor}^\star - \bb' \bSigma \bb)\}_{(s,t) \in \mathcal D} \rightsquigarrow \{\mathbb B(t) - \mathbb B(s)\}_{(s,t) \in \mathcal D},
    \end{equation*}
    where $\mathcal D = \{(s,t),\ 0 \leq s \leq t \leq 1\}$ and $\{\mathbb B(t)\}_{t \in [0,1]}$ is a standard Brownian motion.

\section*{Appendix A.2 Proofs and Lemmas}
In this section, we mainly focus on the mathematical proofs for before the break and when $\epsilon>0$. The results for after the break or $\epsilon<0$ are included, but most of proofs are omitted since they are quite similar. For any fixed $\epsilon \neq 0$, there exists a positive integer $N$ such that when $n \geq N$, $|\epsilon| >(h+1)/n$, therefore, for Lemmas 2-7, we only consider when $|\epsilon|> (|h|+1)/n$. In addition, the model is not distinguishable for all values between $[k/n,(k+1)/n]$ as the break point, so for simplicity we treat $\epsilon n$ as an integer in the proofs. We use $Cs$ to denote generic uniformly positive constants which only depend on the parameters.

\begin{lemma}\label{sigma_x}
For $0 \leq c_1 <c_2 \leq 1$, and $c_1$ and $c_2$ are from the same one of the three intervals, $I_1$, $I_2$ or $I_3$, let
\[
\hat{\bGamma}_x(h, c_1,c_2)=\frac{\sum_{t=1}^{n} \bx_{t} \bx_{t+h}' I_{\{ \lfloor c_1n \rfloor< t \leq \lfloor c_2 n \rfloor \} }}{ N(c_1,c_2)}.
\]
Under Conditions 1 and 2, for any $h \in [1,h_0]$, it holds that
\[
\| \bGamma_x(h,c_1,c_2)\|_2^2\leq k_{\max}^2\sigma_x^4,
\]
\[
\rE\left( \|  \hat{\bGamma}_x(h,c_1,c_2)-\bGamma_x(h,c_1,c_2)\|_2^2\right) \leq \frac{(3h+8\alpha)k_{\max}^2\sigma_x^4}{ N(c_1,c_2)},
\]
where $\alpha=\sum_{t=1}^{\infty} \alpha(t)^{1-2/\zeta}$, and $k_{\max}=\max \{k_1, k_2\}$.
\end{lemma}
\noindent{\it Proof:} Let $a_{q,\ell}$ and $\hat{a}_{q, \ell}$ be the $(q, \ell)$-th entry in $\bGamma_x(h,c_1,c_2)$ and $\hat{\bGamma}_x(h,c_1,c_2)$ respectively. By Condition 2 and Jensen's inequality we know that $\rE(x_{t,i,j}^2)<\sigma_x^2$ and $\rE(x_{t,i,j}^4)<\sigma_x^4$, for $i=1,2$, $j=1,\ldots,k_i$, and $t=1,\ldots,n$. Let $x_{t,q}$ be the $q$-th entry in $\bx_t$. We have $\rE(x_{t,j}^2)<\sigma_x^2$ and $\rE(x_{t,j}^4)<\sigma_x^4$, for $j=1,\ldots, k_1$ when $t\leq \gamma_0 n$, for $j=1,\ldots, k_2$ when $t >\gamma_0 n$.
By Cauchy-Schwarts inequality,
\begin{eqnarray*}
 |a_{q,\ell}|^2 =  \Bigg|\frac{1}{{  N(c_1,c_2)}} \sum_{t=  \lfloor c_1 n \rfloor +1}^{\lfloor c_2 n \rfloor } \rE (x_{t,q}x_{t+h,\ell} )  \Bigg|^2 \leq  \Bigg|\frac{1}{ N(c_1,c_2)}  \sum_{t=  \lfloor c_1 n \rfloor +1}^{\lfloor c_2 n \rfloor }\sqrt{\rE (x_{t,q}^2) \rE(x_{t+h,\ell}^2 )}  \Bigg|^2 =\sigma_x^4.
\end{eqnarray*}
It follows $\|\bGamma_x(h,c_1,c_2)\|_2^2\leq \|\bGamma_x(h,c_1,c_2)\|^2_F\leq k_{\max}^2 \sigma_x^4$.\\
By Proposition 2.5 in \cite{fan2003},
\begin{eqnarray*}
\lefteqn{\rE (\hat{a}_{q,\ell}-a_{q,\ell})^2= \frac{1}{{ N(c_1,c_2)^2}} \rE\Bigg| \sum_{t=1}^{n} [x_{t,q}x_{t+h,\ell}- E (x_{t,q}x_{t+h,\ell})] I_{ \{ \lfloor c_1 n \rfloor <t \leq \lfloor c_2 n \rfloor \}}  \Bigg|^2}\\
&=&  \frac{1}{{ (N(c_1,c_2)^2}} \sum_{\substack{|t_1-t_2|\leq h\\ \lfloor c_1 n \rfloor  < t_1,t_2 \leq \lfloor c_2 n \rfloor }}  \rE [x_{t_1,q}x_{t_1+h,\ell}- E (x_{t_1,q}x_{t_1+h,\ell}) ][x_{t_2,q}x_{t_2+h,\ell}- E (x_{t_2,q}x_{t_2+h,\ell}) ] \\
&&+ \frac{1}{{ N(c_1,c_2)^2}} \sum_{\substack{|t_1-t_2|> h\\ \lfloor c_1 n \rfloor  <t_1,t_2\leq \lfloor c_2 n \rfloor }}   \rE  [x_{t_1,q}x_{t_1+h,\ell}- E (x_{t_1,q}x_{t_1+h,\ell}) ][x_{t_2,q}x_{t_2+h,\ell}- E (x_{t_2,q}x_{t_2+h,\ell}) ]\\
&\leq&\frac{[(2h+1) N(c_1,c_2)-h^2-h]\sigma_x^4}{ N(c_1,c_2)^2}+\frac{(N(c_1,c_2)-h)\sigma_x^{4}}{ N(c_1,c_2)^2}\sum_{u=1}^{N(c_1,c_2)-2h-1} \alpha(u)^{1-2/\zeta}\\
& \leq &\frac{3h N(c_1,c_2)\sigma_x^4}{ N(c_1,c_2)^2}+\frac{(N(c_1,c_2)-h)\alpha \sigma_x^{4}}{ N(c_1,c_2)^2}< \frac{(3h+8\alpha)\sigma_x^{4}}{N(c_1,c_2)}.
\end{eqnarray*}
Thus, $\rE \|\hat{\bGamma}_x(h,c_1,c_2)- {\bGamma}_x(h,c_1,c_2) \|_2^2 \leq \rE \|\hat{\bGamma}_x(h,c_1,c_2)- {\bGamma}_x(h,c_1,c_2)\|_F^2 \leq (3h+8\alpha)k_{\max}^2\sigma_x^4/N(c_1,c_2)$.
\endp

\begin{lemma}\label{Sigma_yhat}
Under Conditions 1-4 and 6, for $\epsilon \in (-\gamma_0, 1-\gamma_0)$ and $|\epsilon|>(h+1)/n$, it holds that
\begin{align*}
\rE\left( \| \hat{\bSigma}_{y,i}(h, \gamma_0+\epsilon) -\bSigma_{y,i}(h, \gamma_0+\epsilon)\|_2^2\right) \leq 144 (3h+8\alpha)a_1^4 k_{\max}^2\sigma_0^4 p^2n^{-1},
\end{align*}
where $\sigma_0=\max\{ \sigma_x, \sigma_{\varepsilon},1\}$, and $a_1>1$ satisfies $\|\bA_i\|_2 \leq a_1 p^{1/2-\delta_i/2}$, for  i=1,2.
\end{lemma}

\noindent{\it Proof:} When $\epsilon>0$,
\begin{eqnarray*}
\lefteqn{\hat{\bSigma}_{y,1}(h, \gamma_0+\epsilon) -\bSigma_{y,1}(h, \gamma_0+\epsilon)}\\
&=&\left[(\gamma_0-\frac{h}{n}) \bA_1 \left( \hat{\bGamma}_{x}(h, 0, \gamma_0-\frac{h}{n}) - \bGamma_{x}(h, 0, \gamma_0-\frac{h}{n})\right)  \bA_1'  \right. \\
&&+ \frac{h}{n} \bA_1 \left( \hat{\bGamma}_{x}(h, \gamma_0-\frac{h}{n}, \gamma_0) - \bGamma_{x}(h, \gamma_0-\frac{h}{n}, \gamma_0)  \right) \bA_2' \\
&&\left. + (\epsilon-\frac{h}{n}) \bA_2 \left( \hat{\bGamma}_{x}(h, \gamma_0, \gamma_0+\epsilon-\frac{h}{n}) - \bGamma_{x}(h, \gamma_0, \gamma_0+\epsilon-\frac{h}{n})   \right) \bA_2' \right] \\
&&+ \frac{1}{n} \sum_{t=1}^{r_0-h}(\bA_1 \bx_{t,1} \bve_{t+h}' +\bve_{t} \bx_{t+h,1}' \bA_1' +\bve_{t} \bve_{t+h}' )+\frac{1}{n}\sum_{t=r_0-h+1}^{r_0} (\bA_1 \bx_{t,1} \bve_{t+h}' +\bve_{t} \bx_{t+h,2}' \bA_2' +\bve_{t} \bve_{t+h}' ) \\
&&+ \frac{1}{n} \sum_{t=r_0+1}^{r_0+\lfloor \epsilon n\rfloor -h}(\bA_2 \bx_{t,2} \bve_{t+h}' +\bve_{t} \bx_{t+h,2}' \bA_2' +\bve_{t} \bve_{t+h}' )  \\
&=&I_1+I_2+I_3+I_4.
\end{eqnarray*}
Condition 4 implies that there exists a positive constant $a_0$ such that $\|\bA_i\|_2 \leq a_0 p^{1/2-\delta_i/2}$ for $i=1,2$. Let $a_1=\max\{a_0,1\}$. By Lemma \ref{sigma_x} and Condition 3, we have
\begin{eqnarray*}
\lefteqn{\rE\|I_1\|_2^2 \leq   3(\gamma_0-\frac{h}{n})^2 \|\bA_1\|_2^4  \cdot \rE\left( \| \hat{\bGamma}_x(h,0,\gamma_0-\frac{h}{n})-\bGamma_x(h,0,\gamma_0-\frac{h}{n}) \|_2^2\right)}\\
&&+ \frac{3h^2}{n^2} \|\bA_1\|_2^2 \cdot \rE\left( \| \hat{\bGamma}_{x}(h, \gamma_0-\frac{h}{n}, \gamma_0) - \bGamma_{x}(h, \gamma_0-\frac{h}{n}, \gamma_0) \|_2^2 \right) \cdot \|\bA_2\|_2^2\\
&&+ 3(\epsilon-\frac{h}{n})^2\|\bA_2\|_2^4 \cdot \rE\left( \| \hat{\bGamma}_{x}(h, \gamma_0, \gamma_0+\epsilon-\frac{h}{n}) - \bGamma_{x}(h, \gamma_0, \gamma_0+\epsilon-\frac{h}{n}) \|_2^2 \right)\\
&\leq&\frac{3(3h+8\alpha) k_{\max}^2 \sigma_x^{4}}{n^2} \left((\gamma_0-\frac{h}{n}) a_1^4p^{2-2\delta_1}+ha_1^4p^{2-\delta_1-\delta_2}n^{-1}+(\epsilon-\frac{h}{n})a_1^4p^{2-2\delta_2}\right)\\
&\leq&\frac{3(3h+8\alpha)a_1^4 k_{\max}^2 \sigma_x^4}{n} \left(\gamma_0 p^{2-2\delta_1}+{h}p^{2-\delta_1-\delta_2}n^{-1}+\epsilon p^{2-2\delta_2}\right).
\end{eqnarray*}

Since $\bx_t$ and $\bve_t$ are independent, 
\begin{eqnarray}
\lefteqn{\rE\Bigg\|\frac{1}{n}\sum_{t=1}^{r_0-h}\bA_1\bx_{t,1} \bve_{t+h}'\Bigg\|_2^2\leq \|\bA_1\|_2^2\cdot \rE \Big\|\frac{1}{n}\sum_{t=1}^{r_0-h}\bx_{t,1} \bve_{t+h}'\Big\|_F^2}\nonumber \\
& \leq& \frac{a_1^2 p^{1-\delta_1}}{n^2} \sum_{i=1}^{k_i}\sum_{j=1}^p \rE \left(  \sum_{t=1}^{r_0-h} x_{t,1,i} \varepsilon_{t+h,j} \right)^2
\leq \frac{a_1^2 p^{1-\delta_1}}{n^2} \sum_{i=1}^{k_i}\sum_{j=1}^p\rE \left(\sum_{t=1}^{r_0-h} x_{t,1,i}^2 \varepsilon_{t+h,j}^2 \right)\nonumber \\
&\leq& \frac{ \gamma_0 a_1^2 k_{\max} \sigma_x^2\sigma_\varepsilon^2 p^{2-\delta_1}}{n}, \label{xe}
\end{eqnarray}
and
\begin{eqnarray}
\rE\Bigg\|\frac{1}{n}\sum_{t=1}^{r_0-h} \bve_t \bx_{t+h,1}'\bA_1' \Bigg\|_2^2\leq \frac{ \gamma_0 a^2_1 k_{\max} \sigma_x^2\sigma_\varepsilon^2 p^{2-\delta_1}}{n},\label{ex}
\end{eqnarray}
where $\varepsilon_{t,j}$ is the $j$-th entry in $\bve_t$. On the other hand,
\begin{eqnarray*}
\rE  \Big\|\frac{1}{n} \sum_{t=1}^{r_0-h} \bve_t \bve_{t+h}' \Big\|_2^2 \leq \rE \Big\| \frac{1}{n} \sum_{t=1}^{r_0-h} \bve_t \bve_{t+h}' \Big\|_F^2
\leq \frac{1}{n^2}\sum_{t=1}^{r_0-h} \sum_{i=1}^p \sum_{j=1}^p  \rE (\varepsilon_{t,i}^2 \varepsilon_{t+h,j}^2)\leq \frac{\gamma_0 \sigma_\varepsilon^4p^2}{n}.
\end{eqnarray*}
Together with (\ref{xe}) and (\ref{ex}) we have
\begin{eqnarray*}
E\|I_2\|_2^2 \leq  \frac{ 3\gamma_0 a_1^2 k_{\max} \sigma_x^2\sigma_\varepsilon^2 p^{1-\delta_1}}{n}+  \frac{ 3\gamma_0 a^2_1 k_{\max} \sigma_x^2\sigma_\varepsilon^2 p^{1-\delta_1}}{n}+ \frac{3\gamma_0 \sigma_\varepsilon^4p^2}{n}\leq \frac{9 \gamma_0 a^2_1 k_{\max} \sigma_0^4p^2}{n},
\end{eqnarray*}
where $\sigma_0=\max\{ \sigma_x, \sigma_{\varepsilon},1\}$. Similarly, we can show that
\begin{eqnarray*}
\rE \| I_3\|_2^2\leq \frac{9 a^2_{1} h k_{\max}\sigma_0^4p^2}{n^2}, \quad \rE\| I_4\|_2 \leq  \frac{9 \epsilon a^2_{1} k_{\max} \sigma_0^4p^2}{n}.
\end{eqnarray*}
Hence,
\begin{eqnarray*}
\lefteqn{\rE\big\| \hat{\bSigma}_{y,1}(h,\gamma_0+\epsilon)-\bSigma_{y,1}(h,\gamma_0+\epsilon)\big\|_2^2\leq \rE(\|I_1\|_2+\|I_2\|_2+\|I_3\|+\|I_4\|_2)^2}\\
&\leq& 4\rE\|I_1\|_2^2+ 4\rE\|I_2\|_2^2+ 4\rE\|I_3\|_2^2+4\rE\|I_4\|_2^2\\
&\leq &{12 (3h+8\alpha)(\gamma_0+\frac{h}{n}+\epsilon)a^4_1 k_{\max}^2 \sigma_x^4p^2}{n^{-1}}+ {36(\gamma_0+\frac{h}{n}+\epsilon)a_1^2 k_{\max} \sigma_0^4p^2}{n^{-1}}\\
&\leq &48(3h +8\alpha)(\gamma_0+\frac{h}{n}+\epsilon)a_1^4 k_{\max}^2\sigma_0^{4}p^2n^{-1} \leq 144 (3h+8\alpha)a_1^4 k_{\max}^2 \sigma_0^4p^2n^{-1}.
\end{eqnarray*}
When $\epsilon<0$, it can be proven in a similar fashion.
\endp

\begin{lemma}\label{sigma_y}
Under Conditions 1-4 and 6, for $\epsilon \in (-\gamma_0, 1-\gamma_0)$ and $|\epsilon|>(h+1)/n$, it holds that
\begin{align*}
\|\bSigma_{y,1}(h,\gamma_0+\epsilon)\|_2\leq \left\{
\begin{array}{ll}
\gamma_0 a^2_1 k_{\max}  \sigma_x^2p^{1-\delta_1},	&\epsilon \in (-\gamma_0,\frac{-(h+1)}{n});\\
 a^2_1 k_{\max} \sigma_x^2(\gamma_0 p^{1-\delta_1}+hp^{1-\delta_1/2-\delta_2/2}n^{-1} +\epsilon p^{1-\delta_2}), &\epsilon \in (\frac{h+1}{n},1-\gamma_0),
 \end{array}\right.
\end{align*}
and
\begin{align*}
\|\bSigma_{y,2}(h,\gamma_0+\epsilon)\|_2\leq \left\{
\begin{array}{ll}
a^2_1 k_{\max} \sigma_x^2[-\epsilon p^{1-\delta_1}+hp^{1-\delta_1/2-\delta_2/2}n^{-1} +(1-\gamma_0)p^{1-\delta_2}], &\epsilon \in (-\gamma_0,\frac{-(h+1)}{n});\\
(1-\gamma_0)a_1^2 k_{\max} \sigma_x^2p^{1-\delta_2},	& \epsilon \in (\frac{h+1}{n},1-\gamma_0).\\
\end{array}\right.
\end{align*}
\end{lemma}

\noindent{\it Proof:} By the definition of $\bSigma_{y,i}(h,\gamma)$ and Lemma \ref{sigma_x},
when $\epsilon >(h+1)/n$, we have
\begin{eqnarray*}
\lefteqn{\big\|\bSigma_{y,1}(h,\gamma_0+\epsilon)\big\|_2
=\frac{1}{n^2} \Bigg\| \sum_{t=1}^{\gamma_0n-h} E(\by_t \by_{t+h}' )  + \sum_{t=\gamma_0n-h+1}^{r_0} E(\by_t \by_{t+h}' ) +  \sum_{t=\gamma_0n+1}^{\gamma_0n+\lfloor \epsilon n \rfloor-h} E(\by_t \by_{t+h}' )  \Bigg\|_2} \\
& \leq& (\gamma_0-\frac{h}{n}) \|\bA_1\|_2^2 \cdot \big\|\bGamma_{x}(h,0, \gamma_0-\frac{h}{n})\big\|_2 +  \frac{h}{n}\|\bA_1\|_2\cdot \|\bA_2\|_2\cdot \big\|\bGamma_{x}(h,\gamma_0-\frac{h}{n}, \gamma_0)\big\|_2 \\
&& +  (\epsilon-\frac{h}{n})\|\bA_2\|_2^2\cdot \big\|\bGamma_{x}(h, \gamma_0, \gamma_0+\epsilon -\frac{h}{n})\big\|_2\\
& \leq&\gamma_0 a^2_1 k_{\max}\sigma_x^2p^{1-\delta_1}+a^2_1 hk_{\max} \sigma_x^2p^{1-\delta_1/2-\delta_2/2}n^{-1} +\epsilon  a^2_1 k_{\max}\sigma_x^2 p^{1-\delta_2},
\end{eqnarray*}
and
\begin{eqnarray*}
\lefteqn{\big\|\bSigma_{y,2}(h,\gamma_0+\epsilon)\big\|_2}\\
&=&\frac{N(1-\frac{h}{n}-\gamma_0-\epsilon)}{n} \big\|\bA_2 \bGamma_x(h, \gamma_0+\epsilon, 1-\frac{h}{n})  \bA_2'  \big\|_2\\
&\leq&
(1-\gamma_0-\epsilon)a^2_1 k_{\max} \sigma_x^2 p^{1-\delta_2}\leq (1-\gamma_0) a_1^2 k_{\max} \sigma_x^2 p^{1-\delta_2}.
\end{eqnarray*}
\endp

\begin{lemma} \label{lemmaB}
Under Conditions 1-4 and 6, for $\epsilon \in  (-\gamma_0, 1-\gamma_0)$and $|\epsilon|>(h+1)/n$, it holds that
\begin{align*}
\|\bB_1' \bSigma_{y,1}(h,\gamma_0+\epsilon)\|_2 \left\{
\begin{array}{ll}
=0, &\epsilon \in (-\gamma_0,\frac{-(h+1)}{n});\\
\leq \epsilon a^2_1 k_{\max} \sigma_x^2  p^{1-\delta_{2}},	&\epsilon \in (\frac{h+1}{n},1-\gamma_0),
\end{array}
\right.
\end{align*}
and
\begin{align*}
\|\bB_2' \bSigma_{y,2}(h,\gamma_0+\epsilon)\|_2 \left\{
\begin{array}{ll}
\leq  a^2_1 k_{\max}  \sigma_x^2 (-\epsilon p^{1-\delta_1}  +h p^{1-\delta_1/2-\delta_2/2}n^{-1}). &\epsilon \in (-\gamma_0,\frac{-(h+1)}{n});\\
=0, & \epsilon \in (\frac{h+1}{n},1-\gamma_0).
\end{array}
\right.
\end{align*}
\end{lemma}

\noindent{\it Proof:}
When $\epsilon>(h+1)/n$, by Lemma \ref{sigma_x}
\begin{eqnarray*}
\lefteqn{\Big\| \bB_1'  \bSigma_{y,1}(h,\gamma_0+\epsilon)\Big\|_2=\frac{1}{n} \Bigg\| \sum_{t=1}^{\lfloor \gamma_0 n+\epsilon n \rfloor-h}\bB_1'\rE \left(\by_t \by_{t+h}' \right)\Bigg\|_2}\\
&=&\frac{1}{n} \Bigg\| \sum_{t=1}^{r_0}\bB_1' \bA_1 \rE \left( \bx_{t,1} \by_{t+h}'  \right) + \sum_{t=r_0+1}^{r_0+ \lfloor \epsilon n \rfloor-h}\bB_1'\bA_2 \rE( \bx_{t,1} \bx_{t+h}' )  \bA_2'\Bigg\|_2\\
& \leq&\frac{N(\gamma_0, \gamma_0+\epsilon-\frac{h}{n})}{n} \|\bB_1\|_2 \cdot \| \bA_2 \bGamma_x(h, \gamma_0, \gamma_0+\epsilon-\frac{h}{n}) \bA_2' \|_2 \leq \epsilon a^2_1 k_{\max} \sigma_x^2 p^{1-\delta_2},
\end{eqnarray*}
and
\begin{eqnarray*}
\Big\| \bB_2'  \bSigma_{y,2}(h,\gamma_0+\epsilon) \Big\|_2=\frac{1}{n} \Bigg\|\sum_{t=\lfloor \gamma_0 n +\epsilon n \rfloor+1}^{n-h}\bB_2' \bA_2{\rm E}(\by_t \by_{t+h}' ) \bA_2'\Bigg\|_2=0.
\end{eqnarray*}
\endp

\begin{lemma}\label{BA}
Under Conditions 4 and 7, it holds that
\begin{align*}
 \|\bB_1' \bA_2\|_2^2 \geq  a_2^2 d^2 \tau p^{1-\delta_2},\quad
 \|\bB_2' \bA_1\|_2^2 \geq a_2^2d^2 \tau p^{1-\delta_1} ,
\end{align*}
where $a_2$ is a positive constant such that $a_2 p^{1/2-\delta_i/2}\leq \|\bA_i\|_{\min}$ for $i=1,2$, and $\tau=\min\{k_2/k_1, k_1/k_2\}$.
\end{lemma}
\noindent{\it Proof:} Note that
\begin{eqnarray*}
\lefteqn{
\tr \left[ \bQ_2'  \left(
\begin{array}{cc}
\bQ_1 &\bB_1
\end{array}
\right) \left(
\begin{array}{c}
\bQ_1'\\
\bB_1'
\end{array}
\right) \bQ_2\right]
=\tr(\bQ_2' \bQ_1\bQ_1' \bQ_2)+\tr(\bQ_2' \bB_1 \bB_1' \bQ_2)}\\
&=k_{\min} \{1-[\cD(\cM(\bQ_1),\cM(\bQ_2))]^2 \}+\tr(\bQ_2' \bB_1 \bB_1' \bQ_2).
\end{eqnarray*}
On the other hand,
\begin{eqnarray*}
\tr \left[ \bQ_2'  \left(
\begin{array}{cc}
\bQ_1 &\bB_1
\end{array}
\right) \left(
\begin{array}{c}
\bQ_1\\
\bB_1
\end{array}
\right) \bQ_2\right]=\tr(\bQ_2' \bQ_2)=k_2.
\end{eqnarray*}
Hence, $\tr(\bQ_2'\bB_1\bB_1'\bQ_2)=k_{\min} [\cD(\cM(\bQ_1),\cM(\bQ_2))]^2+k_2-k_{\min}$. Then we have $\|\bB_1'\bQ_2\|_2^2\geq \tr(\bQ_2'\bB_1\bB_1'\bQ_2)/k_2\geq \tau d^2$. Condition 4 implies that there exists a positive constant $a_2$ such that $\|\bA_2\|_{\min} \geq a_2 p^{1/2-\delta_i/2}$. It follows
\[
\|\bB_1' \bA_2\| \geq a_2d \tau p^{1/2-\delta_2/2}.
\]
\endp

\begin{lemma}\label{Gmin}
Under Conditions 1-8, we have $G(\gamma_0)=0$ and for $\epsilon \in (\eta_1-\gamma_0, \eta_2-\gamma_0)$ and $|\epsilon|>(h+1)/n$,
\begin{align*}
{G}( \gamma_0+\epsilon)
\geq \left\{
\begin{array}{ll}
a_2^2d^2\tau  (a_2^2 u^2_0 \epsilon^2 p^{2-2\delta_1}/2+a_2^2h_1^*  u^2_0\epsilon p^{2-2\delta_1}n^{-1} - a_1^2 h_1^{*2}k_{\max} \sigma^2_xp^{2-\delta_1-\delta_2}n^{-2}), & \epsilon \in (\eta_1-\gamma_0,\frac{-(h+1)}{n});\\
a_2^4d^2\tau u_0^2(\epsilon^2 p^{2-2\delta_2}-2h_2^*p^{2-2\delta_2}n^{-1}),  & \epsilon \in (\frac{h+1}{n},\eta_2-\gamma_0).
\end{array}
\right.
\end{align*}
\end{lemma}
\noindent{\it Proof:} Under Condition 8, by Lemmas \ref{sigma_x} and \ref{BA}, and Theorem 6 in \cite{merikoski2004}, when $\epsilon < -(h+1)/n$,
\begin{eqnarray*}
\lefteqn{G(\gamma_0 + \epsilon) \geq  \| \bB_2' \bM_2(\gamma_0+ \epsilon) \bB_2 \|_2 \geq  \|\bB_2'  \bSigma_{y,2}(h,\gamma_0+ \epsilon)\|_2^2}\\
&\geq &   \| \bB_2' \bA_1\|_2^2 \Big\| \frac{N(\gamma_0+\epsilon,\gamma_0-\frac{h_1^*}{n})}{n}\bGamma_{x}(h_1^*, \gamma_0 + \epsilon , \gamma_0-\frac{h_1^*}{n}) \bA_1'+ \frac{N(\gamma_0-\frac{h_1^*}{n},\gamma_0)}{n}\bGamma_{x}(h_1^*;\gamma_0-\frac{h}{n},\gamma_0) \bA_2'\Big\|_{\min}^2\\
&\geq &  \| \bB_2' \bA_1\|_2^2  \left[-(\epsilon+h_1^*/n)\big\| \bGamma_{x}(h_1^*, \gamma_0 + \epsilon , \gamma_0-\frac{h_1^*}{n}) \bA_1'\|_{\min}-  h \big\|\bGamma_{x}(h_1, \gamma_0 - \frac{h_1^*}{n} , \gamma_0) \bA_1'\big\|_2\right]^2\\
&\geq & \| \bB_2' \bA_1\|_2^2\left[\frac{(\epsilon+h_1^*/n)^2}{2} \big\| \bGamma_{x}(h_1^*, \gamma_0 + \epsilon , \gamma_0-\frac{h_1^*}{n}) \bA_1'\|_{\min}^2-h^2\big\|\bGamma_{x}(h_1, \gamma_0 - \frac{h_1^*}{n} , \gamma_0) \bA_1'\big\|_2^2 \right]\\
&\geq& a_2^2d^2 \tau p^{1-\delta_1} [(\epsilon+h_1^*/n)^2 a_2^2u^2_0p^{1-\delta_1}/2- a_1^2h_1^{*2}k_{\max}\sigma_x^2 p^{1-\delta_2}n^{-2}]\\
& =&a_2^2d^2\tau (a_2^2 u^2_0\epsilon^2 p^{2-2\delta_1}/2+a_2^2 u^2_0 h_1^*\epsilon  p^{2-2\delta_1}n^{-1} - a_1^2h_1^{*2} k_{\max} \sigma_x^2 p^{2-\delta_1-\delta_2}n^{-2});
\end{eqnarray*}
when $\epsilon>(h+1)/n$,
\begin{eqnarray*}
\lefteqn{G(\gamma_0 + \epsilon) \geq  \| \bB_1' \bM_1(\gamma_0+ \epsilon) \bB_1 \|_2 \geq  \|\bB_1'  \bSigma_{y,1}(h,\gamma_0+ \epsilon)\|_2^2}\\
&\geq &  (\epsilon-h/n)^2 \| \bB_1' \bA_2\|_2^2  \cdot \big\| \bGamma_{x}(h_2^*, \gamma_0, \gamma_0+ \epsilon-\frac{h_2^*}{n}) \bA_2'\big\|_{\min}^2\\
&
=&a_2^4d^2\tau u_0^2(\epsilon^2 p^{2-2\delta_2}-2h_2^*p^{2-2\delta_2}n^{-1}).
\end{eqnarray*}
From Lemma \ref{lemmaB}, we have
\[
G(\gamma_0)=0.
\]
\endp

\begin{lemma}\label{B2}
Under Conditions 1-6, if $p^{\delta_{\max}}n^{-1/2}=o(1)$, with true $k_1$ and $k_2$, as $n,p, \to \infty$, we have
\[
\rE \|\hat{\bB}_{i}(\eta_i)-\bB_i(\eta_i)\|_2^2 \leq C p^{2\delta_i}n^{-1}, \mbox { for } i=1,2,
\]
where $C$ is a generic uniformly positive constant which only depends on the parameters.
\end{lemma}

\noindent{\it Proof:} Let $Y_t=x_{t,i,q}x_{t+h,i,\ell} -\rE(x_{t,i,q}x_{t+h,i,\ell})$. Condition 2 indicates that there exists a positive constant $\sigma_y$ such that  $\rE( |Y_t^{2\zeta}|)<\sigma_y^{2\zeta}$. For any $0\leq c_1<c_2\leq 1$, by Cauchy-Schwartz inequality, Proposition 2.5 in \cite{fan2003} and Lemma 1,
\begin{eqnarray*}
\lefteqn{\frac{1}{N(c_1, c_2)^4} \rE \left( \sum_{t=\lfloor c_1 n \rfloor +1}^{\lfloor c_2 n\rfloor}Y_t^4 \right)}\\
&\leq &\frac{1}{N(c_1, c_2)^4}\Bigg[ \sum_{t=\lfloor c_1 n \rfloor +1}^{\lfloor c_1c_2 n\rfloor} \rE(Y_t)^4 +2 \binom 4 1 \sum_{\lfloor c_1 n \rfloor< t_1 <t_2 \leq \lfloor c_2 n\rfloor} \rE(Y_{t_1}^3 Y_{t_2})+ 2 \binom 4 2 \sum_{\lfloor c_1 n \rfloor< t_1 <t_2 \leq \lfloor c_2 n\rfloor} \rE(Y_{t_1}^2 Y_{t_2}^2)\\
&&+\binom 4 2 \binom 2 1 \sum_{\substack{\lfloor c_1 n \rfloor< t_1,t_2,t_3\leq \lfloor c_2 n\rfloor\\ t_1\neq t_2, t_2\neq t_3, t_1\neq t_3}} \rE(Y_{t_1}^2Y_{t_2}Y_{t_3}) +4! \sum_{\lfloor c_1 n \rfloor< t_1 <t_2 <t_3<t_4\leq \lfloor c_2 n\rfloor}  \rE(Y_1Y_2Y_3Y_4)\Bigg]\\
&<&\frac{\sigma_y^4}{N(c_1, c_2)^3}+\frac{10 \sigma_y^4}{N(c_1,c_2)^2}+\frac{12\sigma_y^4}{N(c_1,c_2)}+\frac{24}{N(c_1,c_2)^4} \sum_{\substack{\lfloor c_1 n \rfloor< t_1 <t_2 <t_3\leq \lfloor c_2 n\rfloor \\ t_3-t_2\leq h}}\rE(Y_1Y_2Y_3Y_4)\\
&&+ \frac{24}{N(c_1,c_2)^4} \sum_{\substack{\lfloor c_1 n \rfloor< t_1 <t_2 <t_3\leq \lfloor c_2 n\rfloor \\ t_3-t_2 > h}}\rE(Y_1Y_2Y_3Y_4)\\
&<& \frac{(23+24h) \sigma_y^4}{N(c_1,c_2)}+\frac{24}{N(c_1,c_2)^4} \sum_{\substack{\lfloor c_1 n \rfloor< t_1 <t_2 <t_3\leq \lfloor c_2 n\rfloor \\ t_3-t_2 > h}}[\cov(Y_{t_1}Y_{t_2},Y_{t_3}Y_{t_4}) +\rE(Y_{t_1}Y_{t_2})\rE(Y_{t_3}Y_{t_4})]\\
&<& \frac{47h \sigma_y^4}{N(c_1,c_2)}+\frac{48 \sigma_y^4}{N(c_1,c_2)} \sum_{u=1}^{\lfloor c_2 n \rfloor -2h} \alpha(u)^{1-2\zeta}+\frac{3}{N(c_1,c_2)^4}\left( \sum_{t_1=\lfloor c_1 n \rfloor +1}^{\lfloor c_2 n \rfloor}  \sum_{t_2=\lfloor c_1 n \rfloor +1}^{\lfloor c_2 n \rfloor} |\cov(Y_{t_1}Y_{t_2})| \right)^2\\
&<&\frac{(47h+48\alpha)\sigma_y^4}{N(c_1,c_2)} +\frac{3\sigma_y^4}{N(c_1,c_2)}\left(\sum_{u=1}^{\lfloor c_2 n \rfloor - \lfloor c_1 n\rfloor -2h} \alpha(u)^{1-2\zeta} \right)^2\\
&\leq &\frac{(47h+48\alpha+192 \alpha^2)\sigma_y^4}{N(c_1,c_2)}.
\end{eqnarray*}
It follows
\begin{eqnarray*}
\lefteqn{\rE\|\hat{\bGamma}_x(h,c_1,c_2)-\bGamma_x(h,c_1,c_2 )\|_2^4}\\
&\leq&\rE\|\hat{\bGamma}_x(h,c_1,c_2)-\bGamma_x(h,c_1,c_2 )\|_F^4 \leq \frac{(47h+48\alpha+192 \alpha^2)k_{\max}^2\sigma_y^4}{N(c_1,c_2)}.
\end{eqnarray*}
Thus we have
\begin{eqnarray*}
\lefteqn{\rE \|\hat{\bSigma}_{y,1}(h,\gamma_1)- \bSigma_{y,1}(h,\gamma_1)\|_2^4}\\
&\leq & \frac{16N(0,\gamma_1-\frac{h}{n})^4}{n^4} \|\bA_1\|_2^8 \cdot \rE\big\| \hat{\bGamma}_x(h,0,\gamma_1-\frac{h}{n})- \bGamma_x(h,0,\gamma_1-\frac{h}{n})\big\|^4_2\\
&& +\frac{16}{n^4} \|\bA_1\|_2^4 \cdot \rE\left(\Big\|\sum_{t=1}^{\gamma_1n-h}  \bx_{t,1} \bve_{t+h}' \Big\|_2^4\right)+\frac{16}{n^4} \|\bA_1\|_2^4 \cdot \rE\left(\Big\|\sum_{t=1}^{\gamma_1n-h}  \bve_{t}\bx_{t+h,1}' \Big\|_2^4\right)  +\frac{16}{n^4} \rE\left(\Big\| \sum_{t=1}^{\gamma_1 n -h}\bve_t \bve_{t+h}' \Big\|_2^4
\right)\\
&\leq& \frac{C_1p^{4-4\delta_1}}{n}+ \frac{16C_2 p^{2-2\delta_1}}{n^4} \rE\left( \sum_{t=1}^{\gamma_1 n-h}\sum_{q=1}^{k_1}\sum_{v=1}^p x_{t,1,q}^2 \epsilon_{t+h,v}^2\right)^2+ \frac{16C_2 p^{2-2\delta_1}}{n^4} \rE\left( \sum_{t=1}^{\gamma_1 n-h}\sum_{q=1}^{k_1}\sum_{v=1}^p \epsilon_{t,q}^2 x_{t+h,1,v}^2\right)^2\\
&&+\frac{16C_3}{n^4} \rE\left( \sum_{t=1}^{\gamma_1 n-h}\sum_{q=1}^{p}\sum_{v=1}^p \epsilon_{t,q}^2 \epsilon_{t+h,v}^2\right)^2\\
&\leq&\frac{C_1p^{4-4\delta_1}}{n}+\frac{C_2p^{2-2\delta_1}}{n^2},
\end{eqnarray*}
where $C_1$,$C_2$ and $C_3$ are positive constants and depend only on the parameters.

Hence, with Lemmas \ref{Sigma_yhat} and \ref{sigma_y}
\begin{eqnarray*}
\lefteqn{\rE \| \hat{\bM}_1(\gamma_1)-\bM_1(\gamma_1)\|_2^2}\\
&\leq & h_0 \sum_{h=1}^{h_0}\rE \|\hat{\bSigma}_{y,1}(h,\gamma_1) \hat{\bSigma}_{y,1}(h,\gamma_1)'-\bSigma_{y,1}(h,\gamma_1) \bSigma_{y,1}(h,\gamma_1)' \|_2^2\\
&\leq & 2h_0 \sum_{h=1}^{h_0}\left[ \rE \| \hat{\bSigma}_{y,1}(h,\gamma_1)- \bSigma_{y,1}(h,\gamma_1) \|_2^4 +\| \bSigma_{y,1}(h,\gamma_1)\|_2^2\cdot  \rE \| \hat{\bSigma}_{y,1}(h,\gamma_1)- \bSigma_{y,1}(h,\gamma_1) \|_2^2\right]\\
&\leq&C p^{4-2\delta_1}n^{-1}.
\end{eqnarray*}

Following the proof of Theorem 1 in \cite{lam2011}, we can reach the conclusion.
\endp

\begin{lemma}\label{Ghat}
Under Conditions 1-8, for $\epsilon \in [-\gamma_0, 1-\gamma_0]$, 
it holds that
\begin{align*}
&\rE |\hat{G}(h,\gamma_0+\epsilon)-G(h,\gamma_0+\epsilon)|\\
&\leq \left\{
\begin{array}{ll}
C_1p^2n^{-1}+C_2 \epsilon p^{2-\delta_1} n^{-1/2}+C_3 \epsilon^2 p^{2-2\delta_1+\delta_2} n^{-1/2}, & \epsilon \in(-\gamma_0,-\frac{2}{n});\\
C_1 p^2 n^{-1},	&\epsilon=0; \\
C_1 p^2n^{-1}+C_2 \epsilon p^{2-\delta_2} n^{-1/2}+C_3 \epsilon^2 p^{2+\delta_1-2\delta_2} n^{-1/2}, & \epsilon \in (\frac{2}{n}, 1-\gamma_0).
\end{array}
\right.
\end{align*}
\end{lemma}

\noindent{\it Proof:} By the definition of $\bM_i(\eta_i)$, we can see that ${\cal M}(\bB_i)={\cal M}(\bB_i(\eta_i))$. It implies that there exists an orthogonal $(p-k_i)\times (p-k_i)$ matrix $\bR_i$ such that $\bB_i=\bB_i(\eta_i) \bR_i$.
\[
G(\gamma)= \sum_{i=1}^2\| \bR_i'\bB_i(\eta_i)' \bM_i(\gamma) \bB_i(\eta_i) \bR_i\|_2= \sum_{i=1}^2\| \bB_i(\eta_i)' \bM_i(\gamma) '\bB_i(\eta_i)\|_2.
\]
By the definition of $\hat{G}(\gamma)$ we have,
\begin{eqnarray}
\lefteqn{|\hat{G}(\gamma)-G(\gamma)|}\nonumber \\
 &\leq &\sum_{i=1}^2 \sum_{h=1}^{h_0}  \Big\| \hat{\bB}_i(\eta_i)'\hat{\bSigma}_{y,i}(h,\gamma) \hat{\bSigma}_{y,i}(h,\gamma)' \hat{\bB}_i - \bB_i(\eta_i)' \bSigma_{y,i}(h,\gamma) \bSigma_{y,i}(h,\gamma)'\bB_i(\eta_i)\Big\|_2 \nonumber\\
&\leq& \sum_{i=1}^2\sum_{h=1}^{h_0} \left[ \big\|\hat{\bB}_i(\eta_i)' \hat{\bSigma}_{y,i}(h,\gamma) - \bB_i(\eta_i)' \bSigma_{y,i}(h,\gamma)\big\|_2^2 \right.\nonumber \\
&&+ \left. 2 \big\| \bB_i(\eta_i)' \bSigma_{y,i}(h,\gamma)\big\|_2 \cdot \big\| \hat{\bB}_i(\eta_i)' \hat{\bSigma}_{y,i}(h,\gamma)-\bB_i(\eta_i)' \bSigma_{y,i}(h,\gamma)\big\|_2 \right] \nonumber\\
&\leq & \sum_{i=1}^2\sum_{h=1}^{h_0}  \left[ \left( \| \hat{\bB}_i(\eta_i) \|_2 \cdot \big\| \hat{\bSigma}_{y,i}(h,\gamma)  -{\bSigma}_{y,i}(h,\gamma)\big\|_2 + \|  \hat{\bB}_i(\eta_i)- \bB_i(\eta_i) \big\|_2 \cdot \| {\bSigma}_{y,i}(h,\gamma) \|_2   \right)^2 \right. \nonumber\\
&&+  2 \| \bB_i(\eta_i)'\bSigma_{y,i}(h,\gamma)\|_2 \left( \| \hat{\bB}_i(\eta_i)\|_2 \|\hat{\bSigma}_{y,i}(h,\gamma)- {\bSigma}_{y,i}(h,\gamma) \|_2 + \|\hat{\bB}_i(\eta_i)- \bB_i(\eta_i)\|_2\| {\bSigma}_{y,i}(h,\gamma)   \|_2 \right)  \Bigg] \nonumber\\
&=& \sum_{i=1}^2 \sum_{h=1}^{h_0} L_{i,1}(h,\gamma)+L_{i,2}(h,\gamma). \label{G} 
\end{eqnarray}

By Lemmas \ref{Sigma_yhat}-\ref{lemmaB} and \ref{B2}, 
\begin{eqnarray*}
\rE(L_{1,1}(h,\gamma_0+\epsilon)) \leq \left\{
\begin{array}{ll}
C_1 p^2 n^{-1}, 	& \epsilon \in(-\gamma_0,-\frac{h+1}{n});\\
C_1 p^2 n^{-1} +C_3 \epsilon^2 p^{2+2\delta_1-2\delta_2}n^{-1},	& \epsilon \in (\frac{h+1}{n}, 1-\gamma_0),
\end{array}
\right.
\end{eqnarray*}
\begin{eqnarray*}
\rE(L_{1,2}(h,\gamma_0+\epsilon)) \left\{
\begin{array}{ll}
=0,	& \epsilon \in(-\gamma_0,-\frac{h+1}{n});\\
\leq C_1 \epsilon p^{2-\delta_2} n^{-1/2} +C_3 \epsilon^2 p^{2+\delta_1-2\delta_2}n^{-1/2},	& \epsilon \in (\frac{h+1}{n}, 1-\gamma_0),
\end{array}
\right.
\end{eqnarray*}
\begin{eqnarray*}
\rE(L_{2,1}(h,\gamma_0+\epsilon))\leq \left\{
\begin{array}{ll}
C_1p^2n^{-1} +C_3 \epsilon^2 p^{2-2\delta_1+2\delta_2}n^{-1},	& \epsilon \in(-\gamma_0,-\frac{h+1}{n});\\
C_1 p^2n^{-1}, & \epsilon \in (\frac{h+1}{n}, 1-\gamma_0),
\end{array}
\right.
\end{eqnarray*}

\begin{eqnarray*}
\rE(L_{2,2}(h,\gamma_0+\epsilon)) \left\{
\begin{array}{ll}
\leq C_1 p^{2-\delta_1/2-\delta_2/2} n^{-3/2} +C_2 \epsilon p^{2-\delta_1} n^{-1/2}+ C_3 \epsilon^2 p^{2-2\delta_1+\delta_2}n^{-1/2}, & \epsilon \in(-\gamma_0,-\frac{h+1}{n});\\
=0, & \epsilon \in (\frac{h+1}{n}, 1-\gamma_0).
\end{array}
\right.
\end{eqnarray*}

From (\ref{G}), it follows,
\begin{align*}
&\rE|\hat{G}(\gamma_0+\epsilon)-G(\gamma_0+\epsilon)|\\
&\leq \left\{
\begin{array}{ll}
C_1 p^2n^{-1}+C_2 \epsilon p^{2-\delta_1} n^{-1/2}+C_3 \epsilon^2 p^{2-2\delta_1+\delta_2} n^{-1/2}, & \epsilon \in(-\gamma_0,-\frac{2}{n});\\
C_1 p^2 n^{-1}, &\epsilon =0; \\
C_1 p^2n^{-1}+C_2 \epsilon p^{2-\delta_2} n^{-1/2}+C_3 \epsilon^2 p^{2+\delta_1-2\delta_2} n^{-1/2}, & \epsilon \in (\frac{2}{n}, 1-\gamma_0).
\end{array}
\right.
\end{align*}
\endp

\noindent{\bf Proof of Proposition 1.} We can apply Theorem 1 in \cite{lam2011} to obtain the conclusions.

\noindent{\bf Proof of Theorem 1.}
Since $G(r) \geq 0$ and $G(r_0)=0,$  for any fixed $\epsilon>(h+1)/n$, it follows that
\begin{eqnarray*}
\lefteqn{P(\hat{r}-r_0>\epsilon) =P[\hat{G}(r_0)> \hat{G}(\hat{r}),\hat{r}>r_0+\epsilon]}\\
&= &P\Big[\hat{G}(r_0) -G(r_0)> \hat{G}(\hat{r})-G(\hat{r})+G(\hat{r}),\hat{r}>r_0+\epsilon\Big]\\
&= &P\Big[\hat{G}(r_0) -G(r_0)+G(\hat{r})- \hat{G}(\hat{r})+\frac{3}{4}a_2^4d^2\tau u_0^2\epsilon^2p^{2-2\delta_2}-G(\hat{r})>\frac{3}{4}a_2^4d^2\tau u_0^2\epsilon^2p^{2-2\delta_2},\hat{r}>r_0+\epsilon\Big]\\
&\leq & P\Big[ \big| \hat{G}(r_0)-G(r_0)\big|>+\frac{1}{4}a_2^4d^2\tau u_0^2\epsilon^2p^{2-2\delta_2}\Big] + P\Big[ \big|\hat{G}(\hat{r})-G(\hat{r})\big|>\frac{1}{4}a_2^4d^2\tau u_0^2\epsilon^2p^{2-2\delta_2},\hat{r}>r_0+\epsilon \Big]\\
&&+P\Big[ \frac{3}{4}a_2^4d^2\tau u_0^2\epsilon^2p^{2-2\delta_2}-G(\hat{r})>\frac{1}{4}a_2^4d^2 \tau u_0^2\epsilon^2p^{2-2\delta_2}, \hat{r}>r_0+\epsilon \Big] \\
&=& P\Big[ \big| \hat{G}(r_0)-G(r_0)\big|>\frac{1}{4}a_2^4d^2\tau u_0^2\epsilon^2p^{2-2\delta_2} \Big] + P\Big[ \big|\hat{G}(\hat{r})-G(\hat{r})\big|>\frac{1}{4}a_2^4d^2\tau u_0^2\epsilon^2p^{2-2\delta_2},\hat{r}>r_0+\epsilon \Big]\\
&&+P\Big[G(\hat{r})<\frac{1}{2}a_2^4d^2\tau u_0^2\epsilon^2p^{2-2\delta_2}, \hat{r}>r_0+\epsilon \Big] \\
&=&I_1+I_2+I_3.
\end{eqnarray*}
By Lemma \ref{Gmin}, Lemma \ref{Ghat},  and Chebyshev's inequality, if $p^{\max}n^{-1/2}=o(1)$ and $n$ is large enough, when $\hat{r}>r_0+\epsilon$, we have
\begin{eqnarray*}
I_1<C_1p^{2\delta_2}n^{-1},\quad I_2 <\frac{C_2 p^{\delta_2}n^{-1/2}}{\epsilon}, \quad I_3=0.
\end{eqnarray*}
Hence, there exists a constant $C$ such that
\begin{eqnarray*}
P(\hat{r}>r_0+\epsilon)\leq
\frac{Cp^{\delta_2} n^{-1/2}}{\epsilon}, \mbox{ for } \epsilon >0.
\end{eqnarray*}
\endp

\noindent{\bf Proof of Theorem 2.} When $\hat{\gamma}>\gamma_0$, from Proposition 1, it follows
\[
{\cal D}\{ {\cal M}[\hat{\bQ}_2(\hat{\gamma})], {\cal M}(\bQ_2) \} =O_p(p^{\delta_2} n^{-1/2}), \mbox{ as } n, p \to \infty.
\]
Now we start to investigate the asymptotic properties of $\cM[\hat{\bQ}_1(\hat{\gamma})]$ when $\hat{\gamma}>\gamma_0$.

Lemmas 2-4 imply that
\begin{eqnarray}
\lefteqn{\|\hat{\bM}_1({\gamma_0+\epsilon})-\bM_1({\gamma_0+\epsilon}) \|_2} \nonumber\\
 &\leq & \sum_{h=1}^{h_0}\left( \| \hat{\bSigma}_{y,1}(h,{\gamma_0+\epsilon}) -\bSigma_{y,1}(h,{\gamma_0+\epsilon})\|_2^2 +2 \|\bSigma_{y,1}(h,\gamma_0+\epsilon)\|_2 \cdot \|\hat{\bSigma}_{y,1}(h,{\gamma_0+\epsilon})-\bSigma_{y,1}(h,{\gamma_0+\epsilon})\|_2 \right) \nonumber\\
&=&O_p(p^2 n^{-1})+O_p(p^{2-\delta_1}n^{-1/2})+O_p(\epsilon p^{2-\delta_2} n^{-1/2}) \nonumber\\
&=&O_p(p^{2-\delta_1}n^{-1/2})+O_p(\epsilon p^{2-\delta_2} n^{-1/2}). \label{Ms}
\end{eqnarray}

Under Conditions 2 and 4, it follows from Lemma 1
\begin{eqnarray*}
\lefteqn{\|\bSigma_{y,1}(h,r_0+\epsilon) -\bSigma_{y,1}(h,r_0)\|_2
= \frac{1}{n}  \Big\|\sum_{t=r_0-h+1}^{\gamma_0 +\lfloor \epsilon n \rfloor -h} {\rm E} (\by_t \by_{t+h}') \Big\|_2}\\
&=&\Big\| \frac{N(\gamma_0-h/n,\gamma_0)}{n}\bA_1\bGamma_x(h, \gamma_0-h/n,\gamma_0)\bA_2'+  \frac{N(\gamma_0,\gamma_0+\epsilon-h/n)}{n}\bA_2\bGamma_x(h,\gamma,\gamma_0+ \epsilon -h/n )\bA_2'\Big\|_2\\
&=& O(\epsilon p^{1-\delta_2}) + O(p^{1-\delta_2/2-\delta_{\min}/2} n^{-1}).
\end{eqnarray*}
Hence,
\begin{eqnarray*}
\lefteqn{\|\bM_1({r_0+\epsilon}) -\bM_1\|_2}\\
&\leq &  \sum_{h=1}^{h_0} \big\| \bSigma_{y,1}(h,r_0+\epsilon)\bSigma_{y,1}(h,r_0+\epsilon)' -\bSigma_{y,1}(h,r_0)\bSigma_{y,1}(h,r_0)' \big\|_2\\
&\leq &  \sum_{h=1}^{h_0}  \left(\| \bSigma_{y,1}(h,{r}_0+\epsilon) - \bSigma_{y,1}(h,r_0)\|_2^2+ 2\|\bSigma_{y,1}(h,r_0)\|_2\cdot \|\bSigma_{y,1}(h,r_0+\epsilon)-\bSigma_{y,1}(h,{r_0}) \|_2\right)\\
&=& O(\epsilon^2 p^{2-2\delta_2})+ O(p^{2-\delta_2-\delta_{\min}} n^{-2})+ O(\epsilon p^{2-\delta_1-\delta_2}) + O(p^{2-\delta_1-\delta_2/2-\delta_{\min}/2} n^{-1}).
\end{eqnarray*}
If $p^{\delta_{\max}}n^{-1/2}=o(1)$, together with (\ref{Ms}), we have
\begin{eqnarray*}
\lefteqn{\| \hat{\bM}_1(r_0+\epsilon) -\bM_1\|_2} \nonumber\\
& \leq & \| \hat{\bM}_1(r_0+\epsilon) -{\bM}_1(r_0+\epsilon) \|_2 + \|\bM_1(r_0+\epsilon) -\bM_1\|_2\\
&=&O_p(p^{2-\delta_1}n^{-1/2})+ O(\epsilon p^{2-\delta_1-\delta_2})+O(\epsilon^2 p^{2-2\delta_2}).
\end{eqnarray*}

Theorem 1 tells us if $\hat{r}>r_0$, $|\hat{r}-r_0|=O_p(p^{\delta_{2}} n^{-1/2})$. Therefore,
\begin{eqnarray*}
\| \hat{\bM}_1(\hat{r}) -\bM_1\|_2 =O_p(p^{2-\delta_1} n^{-1/2}).
\end{eqnarray*}
Under Condition 5, by Theorem 9 in \cite{merikoski2004}, we can see that
\begin{eqnarray*}
\|\bM_1\|_{\min} =\|\bSigma_{y,1}(h,\gamma_0)\|^2_{\min}
\geq \|\bA_1\|_2^2 \, \|\bSigma_{x,1}(h,\gamma_0) \|_{\min}^2\, \|\bA_1\|_2^2
=O(p^{2-2\delta_1}).
\end{eqnarray*}
Following the proof of Theorem 2 in \cite{liu2016}, we have
\[
{\cal D}\{ {\cal M}[\hat{\bQ}_1(\hat{\gamma})], {\cal M}(\bQ_1)\}=O_p(p^{\delta_1} n^{-1/2}),
\]
as $n,p \to \infty$, when $\hat{r}>r_0$.

The conclusions for $\hat{r}<r_0$ can be proven in a similar way.
\endp

{\noindent}{\bf Proof of Theorem \ref{thm:cptest}}. The key idea is to show that the test statistic (\ref{eqn:Tn}) can be approximated by a functional of $\{(\lfloor nt \rfloor - \lfloor ns \rfloor \vee 1 + 1) (\hat \nu_{\lfloor ns \rfloor \vee 1,\lfloor nt \rfloor}^\star - \bb' \bSigma \bb)\}_{(s,t)}$ with a negligible difference under the null hypothesis of no change point. For this, by Condition 10 in Appendix A.1, we have
\begin{equation}\label{eqn:proofeq1}
\sup_{s \in (\eta_1,\eta_2)} |\sigma_{n,p}^{-1}\lfloor ns \rfloor (\hat \nu_{1,\lfloor ns \rfloor}^\star - \bb' \bSigma \bb)| \to_D \sup_{s \in (\eta_1,\eta_2)}|\mathbb B(s)|,
\end{equation}
and
\begin{equation}\label{eqn:proofeq2}
\sup_{s \in (\eta_1,\eta_2)} |\sigma_{n,p}^{-1}(n - \lfloor ns \rfloor)(\hat \nu_{\lfloor ns \rfloor+1,n}^\star - \bb' \bSigma \bb)| \to_D \sup_{s \in (\eta_1,\eta_2)}|\mathbb B(1) - \mathbb B(s)|.
\end{equation}
Therefore, the left hand sides of both (\ref{eqn:proofeq1}) and (\ref{eqn:proofeq2}) are of order $O_p(1)$. Let
\begin{equation*}
\phi_{s,n} = (1-s)\lfloor ns \rfloor (\hat \nu_{1,\lfloor ns \rfloor}^\star - \bb' \bSigma \bb) - s(n - \lfloor ns \rfloor)(\hat \nu_{\lfloor ns \rfloor + 1,n}^\star - \bb' \bSigma \bb),
\end{equation*}
then we have
\begin{eqnarray*}
 & & n^{-1}\lfloor ns \rfloor(n-\lfloor ns \rfloor)(\hat \nu_{1,\lfloor ns \rfloor} - \hat \nu_{\lfloor ns \rfloor+1,n}) - \phi_{s,n} \\
 & = & \left(s - {\lfloor ns \rfloor \over n}\right)\{\lfloor ns \rfloor (\hat \nu_{1,\lfloor ns \rfloor}^\star - \bb' \bSigma \bb) + (n - \lfloor ns \rfloor)(\hat \nu_{\lfloor ns \rfloor + 1,n}^\star - \bb' \bSigma \bb)\},
\end{eqnarray*}
and thus by (\ref{eqn:proofeq1}) and (\ref{eqn:proofeq2}),
\begin{equation*}
\sup_{s \in (\eta_1,\eta_2)} |\sigma_{n,p}^{-1}\{n^{-1}\lfloor ns \rfloor(n-\lfloor ns \rfloor)(\hat \nu_{1,\lfloor ns \rfloor} - \hat \nu_{\lfloor ns \rfloor+1,n}) - \phi_{s,n}\}| = O_p(n^{-1}).
\end{equation*}
By condition 10 in Appendix A.1,
\begin{equation*}
\sigma_{n,p}^{-1} \phi_{s,n} \to_D (1 - s) \mathbb B(s) - s \{\mathbb B(1) - \mathbb B(s)\} = \mathbb B(s) - s\mathbb B(1),
\end{equation*}
we have
\begin{equation*}
\sigma_{n,p}^{-1}\{n^{-1}\lfloor ns \rfloor(n-\lfloor ns \rfloor)(\hat \nu_{1,\lfloor ns \rfloor} - \hat \nu_{\lfloor ns \rfloor+1,n}) \to_D \mathbb B(s) - s\mathbb B(1).
\end{equation*}
Note that one can write
\begin{equation*}
{\{r(n-r)(\hat \nu_{1,r} - \hat \nu_{r+1,n})\}^2 \over n^2V_r} = {\{\sigma_{n,p}^{-1} n^{-1} r(n-r)(\hat \nu_{1,r} - \hat \nu_{r+1,n})\}^2 \over \sigma_{n,p}^{-2} V_r},
\end{equation*}
the result follows by the continuous mapping theorem.
\endp

\newpage
\bibliographystyle{asa}
\bibliography{reference1}

\end{document}

%% file: alias.tex
\newtheorem{theorem}{Theorem}
\newtheorem{lemma}{Lemma}

\newtheorem{proposition}{Proposition}

\renewcommand{\hat}{\widehat}
\def\singlespace{\def\baselinestretch{1}\@normalsize}

\def\endp{{\hfill \vrule width 5pt height 5pt\par}}

\newcommand{\cov}{{\rm Cov}}

\newcommand{\tr}{\mbox{tr}}

\newcommand{\bA}{{\mathbf A}}
\newcommand{\bB}{{\mathbf B}}

\newcommand{\bH}{{\mathbf H}}
\newcommand{\bI}{{\mathbf I}}

\newcommand{\bM}{{\mathbf M}}
\newcommand{\bO}{{\mathbf O}}
\newcommand{\bQ}{{\mathbf Q}}

\newcommand{\bR}{{\mathbf R}}
\newcommand{\bS}{{\mathbf S}}
\newcommand{\bU}{{\mathbf U}}

\newcommand{\bb}{{\mathbf b}}

\newcommand{\bq}{{\mathbf q}}

\newcommand{\bu}{{\mathbf u}}
\newcommand{\bv}{{\mathbf v}}

\newcommand{\bx}{{\mathbf x}}
\newcommand{\by}{{\mathbf y}}

\newcommand{\bSigma}{\boldsymbol{\Sigma}}

\newcommand{\bve}{\mbox{\boldmath$\varepsilon$}}

\newcommand{\bGamma} {\boldsymbol{\Gamma}}

\newcommand{\calF}{{\mathcal F}}

\def\6bullets{\bullet\bullet\bullet\bullet\bullet\bullet}


%% file: CPFA.bbl
\begin{thebibliography}{34}
\newcommand{\enquote}[1]{``#1''}
\expandafter\ifx\csname natexlab\endcsname\relax\def\natexlab#1{#1}\fi

\bibitem[{Bai and Ng(2002)}]{bai2002}
Bai, J. and Ng, S. (2002), \enquote{Determining the number of factors in
  approximate factor models.} \textit{Econometrica}, 70, 191--221.

\bibitem[{Bai et~al.(2016)Bai, Taqqu, and Zhang}]{Bai:Taqqu:Zhang:2016}
Bai, S., Taqqu, M.~S., and Zhang, T. (2016), \enquote{A unified approach to
  self-normalized block sampling,} \textit{Stochastic Processes and their
  Applications}, 126, 2465--2493.

\bibitem[{Baltagi et~al.(2017)Baltagi, Kao, and Wang}]{baltagi2017}
Baltagi, B., Kao, C., and Wang, F. (2017), \enquote{Identification and
  estimation of a large factor model with structural instability.}
  \textit{Journal of Econometrics}, 197, 87--100.

\bibitem[{Barigozzi et~al.(2018)Barigozzi, Cho, and Fryzlewicz}]{barigozzi2018}
Barigozzi, M., Cho, H., and Fryzlewicz, P. (2018), \enquote{Simultaneous
  multiple change-point and factor analysis for high-dimensional time series.}
  \textit{Journal of Econometrics}, 206, 187--225.

\bibitem[{Basu and Michailidis(2015)}]{Basu:Michailidis:2015}
Basu, S. and Michailidis, G. (2015), \enquote{Regularized estimation in sparse
  high-dimensional time series models,} \textit{The Annals of Statistics}, 43,
  1535--1567.

\bibitem[{Breitung and Eickmeier(2011)}]{breitung2011}
Breitung, J. and Eickmeier, S. (2011), \enquote{Testing for structural breaks
  in dynamic factor models.} \textit{Journal of Econometrics}, 163, 71--84.

\bibitem[{Chamberlain and Rothschild(1983)}]{chamberlain1983}
Chamberlain, G. and Rothschild, M. (1983), \enquote{Arbitrage, factor structure
  and mean-variance analysis in large asset markets,} \textit{Econometrica},
  70, 191--221.

\bibitem[{Chang et~al.(2015)Chang, Guo, and Yao}]{chang2015}
Chang, J., Guo, B., and Yao, Q. (2015), \enquote{High dimensional stochastic
  regression with latent factors, endogeneity and nonlinearity.}
  \textit{Journal of Econometrics}, 189, 297--312.

\bibitem[{Chang et~al.(2017)Chang, Yao, and Zhou}]{chang2017}
Chang, J., Yao, Q., and Zhou, W. (2017), \enquote{Test for high-dimensional
  white noise using maximum cross correlations.} \textit{Biometrica}, 104,
  1--17.

\bibitem[{Chen(2015)}]{chen2015}
Chen, L. (2015), \enquote{Estimating the common break date in large factor
  models.} \textit{Economics Letters}, 131, 70--74.

\bibitem[{Chen et~al.(2014)Chen, Dolado, and
  Gonzalo}]{Chen:Dolado:Gonzalo:2014}
Chen, L., Dolado, J.~J., and Gonzalo, J. (2014), \enquote{Detecting big
  structural breaks in large facotr models,} \textit{Journal of Econometrics},
  180, 30--48.

\bibitem[{Curry and Shibut(2000)}]{curry2000}
Curry, T. and Shibut, L. (2000), \enquote{The cost of the savings and load
  crisis: truth and consequences.} \textit{FDIC Banking Review.}, 13, 26--35.

\bibitem[{Davis et~al.(2016)Davis, Zang, and Zheng}]{Davis:Zang:Zhen:2016}
Davis, R.~A., Zang, P., and Zheng, T. (2016), \enquote{Sparse vector
  autoregressive modeling,} \textit{Journal of Computational and Graphical
  Statistics}, 25, 1077--1096.

\bibitem[{Doz et~al.(2011)Doz, Giannone, and Reichlin}]{doz2011}
Doz, C., Giannone, D., and Reichlin, L. (2011), \enquote{A two-step estimator
  for large approximate dynamic factor models based on {Kalman} filtering.}
  \textit{Journal of Econometrics}, 164, 188--205.

\bibitem[{Fan and Yao(2003)}]{fan2003}
Fan, J. and Yao, Q. (2003), \textit{Nonlinear Time Series: Nonparametric and
  Parametric Methods}, Springer-Verlag, New York.

\bibitem[{Forni et~al.(2005)Forni, Hallin, Lippi, and Reichlin}]{forni2005}
Forni, M., Hallin, M., Lippi, M., and Reichlin, L. (2005), \enquote{The
  generalized dynamic factor model: one-sided estimation and forecasting.}
  \textit{Journal of the American Statistical Association}, 100, 830--840.

\bibitem[{Forni et~al.(2004)Forni, Lippi, and Reichlin}]{forni2004}
Forni, M., Lippi, M., and Reichlin, L. (2004), \enquote{The generalized dynamic
  factor model: consistency and rates.} \textit{Journal of Econometrics}, 119,
  231--255.

\bibitem[{Fryzlewicz(2014)}]{Fryzlewicz:2014}
Fryzlewicz, P. (2014), \enquote{Wild binary segmentation for multiple
  change-point detection,} \textit{The Annals of Statistics}, 42, 2243--2281.

\bibitem[{Han and Inoue(2015)}]{han2015}
Han, X. and Inoue, A. (2015), \enquote{Tests for parameter instability in
  dynamic factor models.} \textit{Econometric Theory}, 31, 1117--1152.

\bibitem[{Lam and Yao(2012)}]{lam2012}
Lam, C. and Yao, Q. (2012), \enquote{Factor modeling for high-dimensional time
  series: inference for the number of factors,} \textit{Annals of Statistics},
  40, 694--726.

\bibitem[{Lam et~al.(2011)Lam, Yao, and Bathia}]{lam2011}
Lam, C., Yao, Q., and Bathia, N. (2011), \enquote{Estimation of latent factors
  for high-dimensional time series,} \textit{Biometrika}, 98, 901--918.

\bibitem[{Liu and Chen(2019+)}]{liu2019}
Liu, X. and Chen, E. (2019+), \enquote{Helping effects against curse of
  dimensionality in threshold factor models for matrix time series.}
  \textit{Manuscript}, Available at arXiv: 1904.07383.

\bibitem[{Liu and Chen(2016)}]{liu2016}
Liu, X. and Chen, R. (2016), \enquote{Regime-switching factor models for
  high-dimensional time series.} \textit{Statistica Sinica}, 26, 1427--1451.

\bibitem[{Liu and Chen(2019)}]{liu}
--- (2019), \enquote{Threshold factor models for high-dimensional time series.}
  \textit{Manuscript}, Available at arXiv: 1809.03643.

\bibitem[{Lobato(2001)}]{Lobato:2001}
Lobato, I.~N. (2001), \enquote{Testing that a dependent process is
  uncorrelated,} \textit{Journal of the American Statistical Association}, 96,
  1066--1076.

\bibitem[{Ma and Su(2018)}]{Ma:Su:2018}
Ma, S. and Su, L. (2018), \enquote{Estimation of large dimensional factor
  models with an unknown number of breaks,} \textit{Journal of Econometrics},
  207, 1--29.

\bibitem[{Merikoski and Kumar(2004)}]{merikoski2004}
Merikoski, J.~K. and Kumar, R. (2004), \enquote{Inequalities for spreads of
  matrix sums and products.} \textit{Applied Mathematics E-Notes}, 4, 150--159.

\bibitem[{Shao(2010)}]{Shao:2010}
Shao, X. (2010), \enquote{A self-normalized approach to confidence interval
  construction in time series,} \textit{Journal of the Royal Statistical
  Society: Series B}, 72, 343--366.

\bibitem[{Shao(2011)}]{Shao:2011:CPLRD}
--- (2011), \enquote{A simple test of changes in mean in the possible presence
  of long-range dependence,} \textit{Journal of Time Series Analysis}, 32,
  598--606.

\bibitem[{Stock and Watson(1998)}]{stock1998}
Stock, J.~H. and Watson, M.~W. (1998), \enquote{Diffusion indexes.}
  \textit{NBER Working Paper 6702}.

\bibitem[{Stock and Watson(2005)}]{stock2005}
--- (2005), \enquote{Implications of dynamic factor models for {VAR} analysis.}
  \textit{National Bureau of Economic Research}, Working Paper 11467.

\bibitem[{Taqqu and Zhang(2019)}]{Taqqu:Zhang:2019}
Taqqu, M.~S. and Zhang, T. (2019), \enquote{A self-normalized semiparametric
  test to detect changes in the long memory parameter,} \textit{Journal of Time
  Series Analysis}, forthcoming.

\bibitem[{Van~der Vaart and Wellner(1996)}]{vanderVaart:Wellner:1996}
Van~der Vaart, A.~W. and Wellner, J.~A. (1996), \textit{Weak Convergence and
  Empirical Processes}, New York: Springer Verlag.

\bibitem[{Wang et~al.(2019)Wang, Liu, and Chen}]{wang2019}
Wang, D., Liu, X., and Chen, R. (2019), \enquote{Factor Models for
  Matrix-Valued High-Dimensional Time Series,} \textit{Journal of
  Econometrics}, 208, 231--248.

\end{thebibliography}
